\input{aipcheck}

\documentclass[
    ,final            
  ,numberedheadings 
  ]
  {aipproc}

\layoutstyle{6x9}

\usepackage{amsmath,amssymb}
\usepackage{color}

\begin{document}

\title{The von Neumann Model of Measurement in Quantum Mechanics}


\author{Pier A. Mello}{
  address={Instituto de F\'isica, Universidad Nacional Aut\'onoma de M\'exico,
Apdo. Postal 20-364, 01000 M\'exico, D. F., M\'exico}
}
\begin{abstract}

We describe how to obtain information on a quantum-mechanical system by coupling it to a probe and detecting some property of the latter, using a model introduced by von Neumann, which describes the interaction of the system proper with the probe in a dynamical way.

We first discuss single measurements, where the system proper is coupled to one probe with arbitrary coupling strength.
The goal is to obtain information on the system detecting the probe position.
We find the reduced density operator of the system, and show how L\"uders rule emerges as the limiting case of strong coupling.

The von Neumann model is then generalized to two probes that interact successively with the system proper.
Now we find information on the system by detecting the position-position and momentum-position correlations of the two probes.
The so-called ``Wigner's formula" emerges in the strong-coupling limit, while ``Kirkwood's quasi-probability distribution" is found as the weak-coupling limit of the above formalism.
We show that successive measurements can be used to develop a state-reconstruction scheme.

Finally, we find a generalized transform of the state and the observables based on the notion of successive measurements.

\end{abstract}

\maketitle

\section{INTRODUCTION}

In a general quantum measurement one obtains information on the system of interest
by coupling it to an {\em auxiliary degree of freedom}, or {\em probe}, 
and then {\em detecting some property of the latter} using a measuring device.
This procedure, which was described by von Neumann in his classic book 
\cite{Neumann-MathFounQuanMech:55}, will be referred to as von Neumannn's model (vNM).
Within the vNM, the combined system --system proper plus probe-- is given a 
dynamical description.

An example of the idea involved in the vNM is given by the Stern-Gerlach experiment, which is described in every textbooks on QM 
(see, e.g., \cite{ballentine:99,peres-book}).

\begin{figure}[h]
\includegraphics[height=.2\textheight]{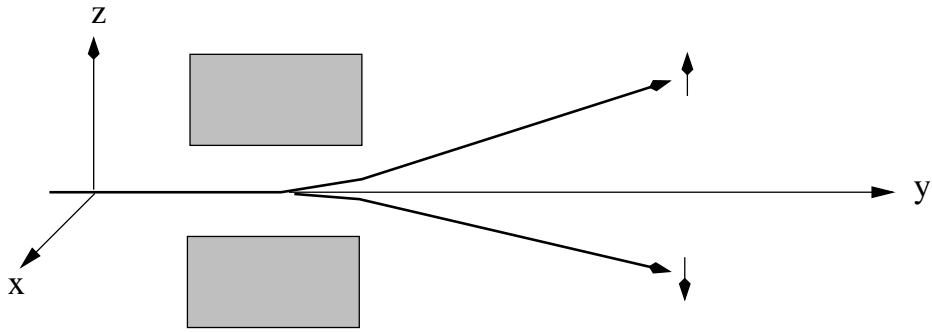}
  \caption{The Stern-Gerlach experiment, designed to find information on the 
  $z$-component of the spin of a particle by detecting its position.}
\label{fig1}
\end{figure}
In this experiment, presented schematically in Fig. \ref{fig1}, the observable we want to obtain information about is
the $z$-component of the spin of a particle.
The auxiliary degree of freedom, or probe, is its position ${\bf r}$
--still a microscopic quantity-- after it leaves the magnet, and this is what is 
recorded by a detecting device, which in this case is a  position detector.
E.g., for $s=1/2$, we may wish to find information on the two components  
$\langle \psi_{spin}|\mathbb{P}^z_{\pm}| \psi_{spin}\rangle =|\langle\pm |\psi_{spin}\rangle|^2$ of the original state.

Another example is provided by Cavity Quantum Electrodynamics (QED) experiments
\cite{brune-et-al-1992,davidovich-et-al-1996,guerlin_et_al_2007}.
\begin{figure}[h]
\includegraphics[height=.2\textheight]{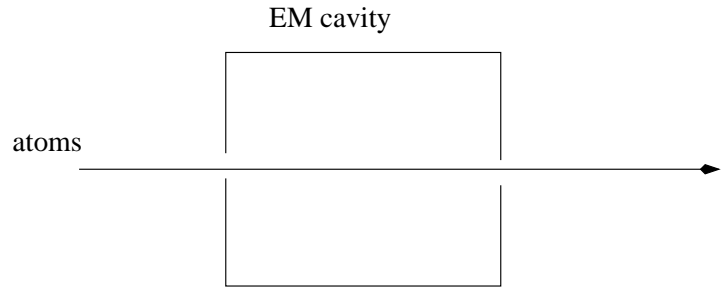}
\caption{A Cavity QED experiment, designed to find information on the 
number of photons inside a cavity, by sending atoms through the cavity 
and subsequently detecting them.}
\label{fig2}
\end{figure}
Here the observable is the number $n$ of photons in a cavity. 
As shown schematically in Fig. \ref{fig2}, the probes are atoms successively sent through the cavity: after they leave the cavity, they are detected by a measuring device. 

In what follows we first discuss, in the next section, the Stern-Gerlach problem, which has become a paradigm for models of measurement in QM.

The discussion will pave the way for the analysis, in Sec. \ref{single}, of single measurements in QM, where the system proper is coupled to one probe with arbitrary coupling strength.
The main goal is to study what information can we obtain on the system by detecting the probe position \cite{johansen-mello:2008}.
As a by-product of the analysis, we obtain the reduced density operator of the system proper after its interaction with the probe, and derive the so-called L\"uders rule
\cite{Lueders-UEbeZust:51} as the limiting case of strong coupling.

We then generalize the vNM to two probes that interact successively with the system proper \cite{johansen-mello:2008}.
Again, we study what information can we obtain on the system by detecting the position-position and momentum-position correlations of the two probes.
Indeed, we describe a state reconstruction scheme based on the procedure of successive measurements \cite{johansen-mello:2008,amir-pier}.
We obtain the so-called ``Wigner's formula" 
\cite{Wigner-ProbMeas:63}
in the strong-coupling limit of the above formalism, and
Kirkwood's quasi-probability distribution 
\cite{Kirkwood-QuanStatAlmoClas:33}
in the weak-coupling limit.
We also find a generalized transform of the state and the observables based on the notion of successive measurements, and in terms of complex quasi-probabilities.

\section{The Stern-Gerlach experiment}
\label{stern-gerlach}

The Stern-Gerlach experiment, Fig. \ref{fig1}, has become a paradigm for models of measurement in QM.
Although the experiment was first performed as early as 1922, 
it is explained in every textbook on QM 
(see, e.g. Refs. \cite{ballentine:99,peres-book}),
and various refinements have been presented in the literature           (as in Refs. \cite{peres-1980,alstrom-1982,platt-1992}),
surprisingly, its complete (non-relativistic) solution has been given only recently \cite{scully-1987}.

By a complete solution we mean one that takes into account
the {\em translational} and {\em transverse} motions of the atom, and 
a {\em confined} magnetic field ${\bf B}({\bf r})$ that satisfies Maxwell's equations
\begin{equation}
{\bf \nabla \cdot B =0}, \hspace{1cm} {\bf \nabla \times B =0} .
\label{divB=0,rotB=0}
\end{equation}

Here we shall work with a model of the complete problem which has a simple exact solution and still shows the physical characteristic we want to exhibit,
i.e., bending of the trajectory depending on the $z$-projection of the spin.

We shall:
i) assume ${\bf B}\equiv 0$ outside the gaps; ii) assume that only $B_z$ is significant;
iii) assume $B_z(z) \approx B'(z)z$ inside the gaps;
iv) simulate the translational motion in the $y$ direction using a $t$-dependent interaction which lasts for the time the particle is inside the gap of the magnet. 
This could be achieved by adopting a frame of reference moving with the particle;
v) ignore the $x$ degree of freedom.

We then consider the model Hamiltonian
\begin{subequations}
\begin{eqnarray}
\hat{H}(t) &=& \frac{\hat{p}_z^2}{2m} -{\bf \mu}{\bf \cdot B} \; \theta_{t_1,\tau}(t) \\
&=&  \frac{\hat{p}_z^2}{2m} - \left[\mu_B B'_z(0) \tau\right]\frac{\theta_{t_1,\tau}(t)}{\tau} \hat{\sigma}_z \; \hat{z} \;, \\
&=&  \frac{\hat{p}_z^2}{2m} - \epsilon g(t) \hat{\sigma}_z \; \hat{z}\; ,
\end{eqnarray}
\label{H(t) 1} 
\end{subequations}
where $\hat{\sigma}_z$ is one of the Pauli matrices, and $\epsilon =  \mu_B B'_z(0) \tau$.
The function $\theta_{t_1,\tau}(t)$ is nonzero and equal to unity only inside the time interval $(t_1-\tau/2, t_1+\tau/2)$, i.e.,
\begin{eqnarray}
\theta_{t_1,\tau}(t) &=&  
\left\{
\begin{array}{cc}
1, & t \in (t_1-\tau/2, t_1+\tau/2) \\
0 & t  \notin (t_1-\tau/2, t_1+\tau/2)
\end{array}
\right .
\label{theta}
\end{eqnarray}
and
\begin{equation}
g(t)=  \frac{\theta_{t_1,\tau}(t)}{\tau} \; ,
\hspace{5mm}
\int_{0}^{\infty} g(t)dt =1, \;\;\;  (\tau \ll t_1) .
\label{g(t)}
\end{equation}
We further use the simplification $g(t) \approx \delta(t-t_1)$ and write our model Hamiltonian as
\begin{subequations}
\begin{eqnarray}
\hat{H}(t) &=& \frac{\hat{p}_z^2}{2m} -\epsilon \delta(t-t_1) \hat{\sigma}_z \hat{z}
\\
&=&\hat{H}_0 + \hat{V}(t) .
\label{H(t) 2} \\
\hat{H}_0 &=& \hat{K}_z \;\;\;\;
{\rm is \; the \; kinetic \;  energy \;  operator \; for \; the}\; 
z \; {\rm variable}.
\label{H0=Kz}
\end{eqnarray}
\label{H(t),H0}
\end{subequations}
In the nomenclature introduced in the Introduction,
$\hat{\sigma}_z$ is the {\em observable for the system proper} and
$\hat{z}, \hat{p}_z$ are the {\em probe canonical variables}.
From now on we adopt the nomenclature that we {\em measure} 
(perhaps a better word could be ``premeasure") the observable $\hat{\sigma}_z$, by
{\em detecting}, with a suitable instrument, either $\hat{z}$ or $\hat{p}_z$.

We shall solve the Schr\"odinger equation 
\begin{subequations}
\begin{eqnarray}
i\hbar\frac{\partial |\psi(t)\rangle}{\partial t}
&=& \hat{H}(t) |\psi(t)\rangle \; ,
\label{schr. eqn SG}
\end{eqnarray}
with the initial condition
\begin{eqnarray}
|\psi(0)\rangle &=& |\psi_{spin}^{(0)}\rangle  |\chi^{(0)}\rangle \; .
\label{initial condition SG}
\end{eqnarray}
\end{subequations}
Here, $|\psi_{spin}^{(0)}\rangle$ is the initial state of the system proper and
$|\chi^{(0)}\rangle$ the initial state of the probe.

It will be advantageous to use the interaction picture 
(for a textbook presentation, see, e.g., Ref. \cite{messiah}),
in which we have the following relations
\begin{subequations}
\begin{eqnarray}
|\psi(t) \rangle_I &=& \hat{U}_0^{\dagger}(t)|\psi(t) \rangle
\label{inter picture a} \\
&=& \hat{U}_0^{\dagger}(t)\hat{U}(t)|\psi(0) \rangle
\label{inter picture b} \\
&\equiv& \hat{U}_I (t) |\psi(0) \rangle 
\label{inter picture c} \\
\hat{U}_I (t)&=& \hat{U}_0^{\dagger}(t)\hat{U}(t)
\label{inter picture d} \\
i\hbar \frac{d \hat{U}_I (t)}{dt} 
&=& \hat{V}_I(t)\hat{U}_I (t)\; ; \;\;\;\; \hat{U}_I (0)=\hat{I} \; .
\label{inter picture e}
\end{eqnarray}
Here, $\hat{U}_0(t)$, $\hat{U}(t)$,  are the evolution operators in the Schr\"odinger picture associated with $H_0$ and $\hat{H}$, respectively, and $\hat{U}_I(t)$ the evolution operator in the interaction picture; $\hat{V}_I(t)$ is the interaction in the interaction picture, i.e.,
\begin{eqnarray}
\hat{V}_I(t)&=& {\rm e}^{\frac{i}{\hbar}\hat{H}_0 t}\hat{V}(t)
{\rm e}^{-\frac{i}{\hbar}\hat{H}_0 t}
\label{inter picture e} \\
&=& -\epsilon \delta (t-t_1){\rm e}^{\frac{i}{\hbar}\hat{H}_0 t_1} 
\hat{\sigma}_z \hat{z} 
{\rm e}^{-\frac{i}{\hbar}\hat{H}_0 t_1} 
\label{inter picture f} \\
&\equiv& -\epsilon \delta (t-t_1) \hat{W} \; .
\end{eqnarray}
\label{inter picture}
\end{subequations}
The solution for $\hat{U}_I(t)$ is
\begin{subequations}
\begin{eqnarray}
\hat{U}_I(t) &=& {\rm e}^{\frac{i\epsilon}{\hbar}\int_0^t{\delta(t'-t_1)dt' \cdot \hat{W}}}
\label{UI(t)a} \\
\hat{U}_{I,f}
&=& {\rm e}^{\frac{i}{\hbar}\hat{H}_0 t_1}
{\rm e}^{\frac{i\epsilon}{\hbar}\hat{\sigma}_z \hat{z}}
{\rm e}^{-\frac{i}{\hbar}\hat{H}_0 t_1} \; .
\end{eqnarray}
\label{UI(t)}
\end{subequations}
In the last line we have the ``final" evolution operator in the interaction picture, i.e., after the interaction has ceased to act.
From Eq. (\ref{inter picture d}) we find the final evolution operator in the Schr\"odinger picture as
\begin{equation}
U_f \equiv U(t>t_1) 
= e^{-\frac{i}{\hbar}\hat{K}_z(t-t_1)}
e^{\frac{i}{\hbar}\epsilon \hat{\sigma}_z \hat{z}}
e^{-\frac{i}{\hbar}\hat{K}_z t_1} \; ,
\label{Uf}
\end{equation}
whose physical interpretation is very clear: we first have free evolution from $t=0$ to $t_1$; the interaction acts at $t=t_1$, and we have free evolution again thereafter.

The final state vector, i.e., for $t>t_1$, is then
\begin{equation}
|\Psi(t)\rangle_f
= \sum_{\sigma = \pm 1} \mathbb{P}_{\sigma}|\psi_{spin}^{(0)}\rangle 
e^{-\frac{i}{\hbar}\hat{K}_z(t-t_1)}
e^{\frac{i}{\hbar}\epsilon \sigma \hat{z}}
|\chi(t_1)\rangle
\label{Psif}
\end{equation}
We have introduced the projector $\mathbb{P}_{\sigma}$ onto the state with eigenvalue  $\sigma=+1,-1$ of $\hat{\sigma}_z$.
We observe that the component 
$\mathbb{P}_{\sigma}|\psi_{spin}^{(0)}\rangle$
of the original spin state gets entangled with
the probe state 
$e^{\frac{i}{\hbar}\epsilon \sigma \hat{z}}|\chi(t_1)\rangle$
(we denote by $|\chi(t_1)\rangle$ the probe state which has evolved freely from $t=0$ to $t=t_1^{-}$)
which, in the $z$-representation, is
\begin{equation}
e^{\frac{i}{\hbar}\epsilon \sigma z}\chi(z,t_1) \; ,
\label{z-boost}
\end{equation}
meaning that it got a boost $p_{\sigma}=\epsilon \sigma$ in the $z$ direction.
This is precisely the essential physical effect occurring in the Stern-Gerlach experiment.

\subsection{The probability of detecting a probe position $z$ as a function of time}
\label{pf(z,t)}

\begin{figure}[h]
\includegraphics[height=.25\textheight]{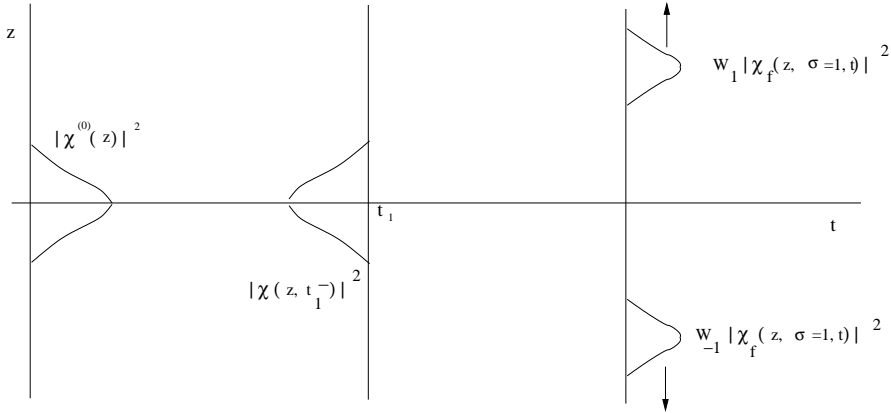}
\caption{The probability $p(z,t)$ of a probe position $z$ as a function of time in the Stern-Gerlach experiment, as explained in the text.
}
\label{fig3}
\end{figure}

At $t=0$, the probability density $p(z,t)$ of a probe position $z$ is just 
$|\chi^{(0)}(z)|^2$, where $\chi^{(0)}(z)$ is the original $z$-wave function.
As time goes on, the wave packet spreads in time until just before the interaction occurs, i.e., until $t=t_1^{-}$ (see the schematic illustration in 
Fig. \ref{fig3}).
After the interaction has taken place at $t=t_1$, i.e, for $t>t_1$, that probability is given by
\begin{subequations}
\begin{eqnarray}
p_f(z,t)
&=& \; _f \langle \Psi| \mathbb{P}_z  |\Psi \rangle_f 
\label{pf(z,t) a}   \\
&=& \sum_{\sigma = \pm 1} 
\langle \psi_{spin}^{(0)}|\mathbb{P}_{\sigma}|\psi_{spin}^{(0)}\rangle 
\left|\langle z |e^{-\frac{i}{\hbar}\hat{K}_z(t-t_1)}
e^{\frac{i}{\hbar}\epsilon \sigma \hat{z}}
|\chi(t_1)\rangle \right|^2 
\label{pf(z,t) b}   \\
&=& 
\sum_{\sigma = \pm 1} W_{\sigma} ^{(\hat{\sigma}_z)}
|\chi_f(z,\sigma;t)|^2 \; ,
\label{pf(z,t) c}
\end{eqnarray}
\label{pf(z,t)}
\end{subequations}
where $W_{\sigma} ^{(\hat{\sigma}_z)}
=\langle \psi_{spin}^{(0)}|\mathbb{P}_{\sigma}|\psi_{spin}^{(0)}\rangle$
is the Born probability for the value $\sigma$ of the spin projection in the original system state.
We have also defined the $\sigma$-dependent probe wave function for $t>t_1$
\begin{subequations}
\begin{eqnarray}
\chi_f(z,\sigma;t) 
&=&\langle z |e^{-\frac{i}{\hbar}\hat{K}_z(t-t_1)}
e^{\frac{i}{\hbar}\epsilon \sigma \hat{z}}
|\chi(t_1)\rangle
\\
&\equiv& \int U_0(z,z';t-t_1) 
e^{\frac{i}{\hbar}\epsilon \sigma z'}
\chi(z', t_1) dz'
\end{eqnarray}
\label{chi-f(z,sigma,t)}
\end{subequations}
which consists of the probe wave function which evolves freely up to $t=t_1$, it gets multiplied by the plane wave $e^{\frac{i}{\hbar}\epsilon \sigma z'}$ at $t=t_1$
(the boost referred to above), and evolves freely thereafter, through the free evolution operator indicated as $U_0$.
Thus, because of the interaction occurring at $t=t_1$, the $\sigma=1$ component of this state receives a positive boost in the $z$ direction, and the $\sigma=-1$ component receives a negative boost, so that after $t=t_1$ they
travel in opposite directions.
From Eq. (\ref{pf(z,t) c}), the $\sigma=1$ component occurs with a weight $W_{1} ^{(\hat{\sigma}_z)}$, and 
the $\sigma=-1$ component with a weight $W_{-1} ^{(\hat{\sigma}_z)}$, as is also indicated in Fig. \ref{fig3}.

\subsubsection{Conditions generally required \cite{brune-et-al-1992,imoto-et-al}
for the measurement of the observable of a system, when detecting a property of the probe
}

What are called $\hat{A}_s$ and $\hat{A}_{probe}$ in 
Refs. \cite{brune-et-al-1992,imoto-et-al},
are translated as
\begin{subequations}
\begin{eqnarray}
\hat{A}_s = \hat{\sigma}_z 
\label{} \\
\hat{A}_{probe}= \hat{z} \; ,
\label{} 
\end{eqnarray}
\label{}
\end{subequations}
respectively, in our notation for the present Stern-Gerlach problem.
The requirements established in these references are:

i) 
\begin{equation}
\hat{V} = f(\hat{A}_s)
\label{}
\end{equation}
Here, indeed: $\hat{V}=-\epsilon \delta(t-t_1) \hat{\sigma}_z \hat{z}$
 
ii) 
\begin{equation}
[\hat{V},\hat{A}_{probe}] \neq 0 \; ,
\label{}
\end{equation}
so that $\langle \hat{A}_{probe} \rangle$ changes,
enabling one to get information on $\hat{A}_s$ by detecting $\hat{A}_{probe}$.
Although here $[\hat{V}, \hat{z}] =0$, 
we have $[\hat{K_z}, \hat{z}] \neq 0$; 
the kinetic energy $\hat{K}_z$ causes a displacement of the two wave packets, as illustrated in Fig. \ref{fig3}:
if we wait long enough, the 2 packets separate, and we can measure 
$\sigma = \pm 1$.

In order to have a
{\em a QM non-demolition measurement}, these references also establish the additional conditions:

iii) 
\begin{subequations}
\begin{eqnarray}
\left[\hat{V},\hat{A}_s \right] &=& 0 \; ;
\;\;\;\;\; {\rm here, \; indeed,} \; \left[\hat{V},\hat{\sigma}_z \right] = 0
\label{} \\
\left[ \hat{H}_s,\hat{A}_s \right] &=& 0 \; ;
\;\;\;\;\; {\rm here},\;\; H_s=0, \;\;\; {\rm so \; that} \;\; [\hat{H}_s,\hat{\sigma}_z ]=0
\label{}
\end{eqnarray}
which, taken together, give
\begin{eqnarray}
\left[ \hat{H},\hat{A}_s \right] &=& 0 \; ;
\;\;\;{\rm here,\; indeed}, \; \left[\hat{H},\hat{\sigma}_z \right] = 0 ,
\end{eqnarray}
\label{}
\end{subequations}
with the consequence that starting with 
$|\psi_{spin}^{(0)} \rangle = |\sigma = 1\rangle$, say, the state will not get a component $|\sigma = -1\rangle$.

\subsection{The probability of detecting a probe momentum $p_z$ as a function of time}
\label{pf(pz,t)}

\begin{figure}[h]
\includegraphics[height=.25\textheight]{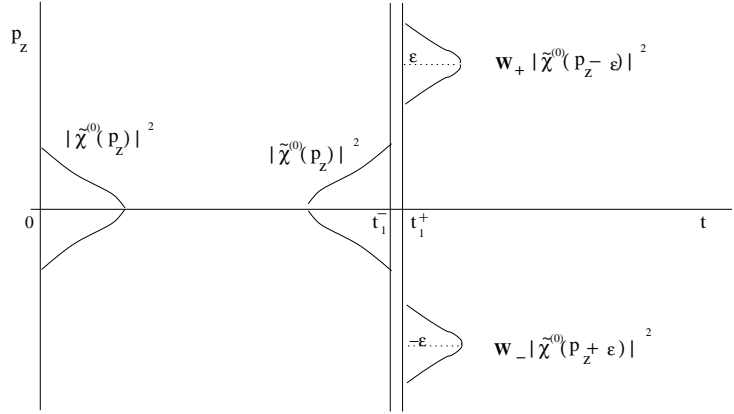}
\caption{The probability $p(z,t)$ of a probe momentum $p_z$ as a function of time in the Stern-Gerlach experiment, as explained in the text.
}
\label{fig4}
\end{figure}

At $t=0$, the probability density $p(p_z,t)$ of a probe momentum $p_z$ is 
$|\tilde{\chi}^{(0)}(p_z)|^2$, where $\tilde{\chi}^{(0)}(p_z)$, the original wave function in the momentum representation, is the Fourier transform of the original wave function in $z$-space.
As time goes on, the wave packet conserves its shape until just before the interaction occurs, i.e., until $t=t_1^{-}$ (see the schematic illustration in 
Fig. \ref{fig4}).

After the interaction has taken place at $t=t_1$, i.e, for $t>t_1$, that probability is given by
\begin{subequations}
\begin{eqnarray}
p_f(p_z,t)
&=& \; _f \langle \Psi| \mathbb{P}_{p_z}  |\Psi \rangle_f \\
&=& \sum_{\sigma = \pm 1} 
\langle \psi_{spin}^{(0)}|\mathbb{P}_{\sigma}|\psi_{spin}^{(0)}\rangle 
\left|\langle p_z |e^{-\frac{i}{\hbar}\hat{K}_z(t-t_1)}
e^{\frac{i}{\hbar}\epsilon \sigma \hat{z}}
|\chi(t_1)\rangle \right|^2 \\
&=&
\sum_{\sigma = \pm 1} W_{\sigma} ^{(\hat{\sigma}_z)}
\left|\tilde{\chi}^{(0)}(p_z- \epsilon\sigma)\right|^2 \; .
\end{eqnarray}
\end{subequations}
Just as in Eq. (\ref{pf(z,t)}), $W_{\sigma} ^{(\hat{\sigma}_z)}$
is the Born probability for the value $\sigma$ of the spin projection in the original system state;
$\tilde{\chi}^{(0)}(p_z- \epsilon\sigma)$ is the original wave function in momentum space, evaluated at the displaced value $p_z- \epsilon\sigma$.
Thus, at $t=t_1^+$ the $p_z$ probability density splits into the two pieces indicated in Fig. \ref{fig4}, corresponding to the two values of $\sigma$, with weights $W_{+1} ^{(\hat{\sigma}_z)}$ and $W_{-1} ^{(\hat{\sigma}_z)}$, respectively, and remains unaltered thereafter.

\subsubsection{Conditions generally required \cite{brune-et-al-1992,imoto-et-al}
for the measurement of the observable of a system, when detecting a property of the probe
}

In this case, the $\hat{A}_s$ and $\hat{A}_{probe}$ of
Refs. \cite{brune-et-al-1992,imoto-et-al} are 
\begin{subequations}
\begin{eqnarray}
\hat{A}_s &=& \hat{\sigma}_z 
\label{} \\
\hat{A}_{probe}&=& \hat{p}_z \; ,
\label{} 
\end{eqnarray}
\label{}
\end{subequations}
respectively, in our present notation.
We recall that these references require: 

i) 
\begin{equation}
\hat{V} = f(\hat{A}_s) \; ;
\label{}
\end{equation}
indeed, this is the case, because
$\hat{V}=-\epsilon \delta(t-t_1) \hat{\sigma}_z \hat{z}$.
 
ii) 
\begin{equation}
[\hat{V},\hat{A}_{probe}] \neq 0 \; .
\label{}
\end{equation}
Here, indeed, $[\hat{V}, \hat{p}_z] \propto [\hat{z},\hat{p}_z] \neq0$.

For {\em a QM non-demolition measurement}, these references require:

iii) 
\begin{subequations}
\begin{eqnarray}
\left[\hat{V},\hat{A}_s \right] &=& 0 \; ; 
\;\;\;{\rm here, \; indeed:} \; \left[\hat{V},\hat{\sigma}_z \right] = 0
\label{} \\
\left[ \hat{H}_s,\hat{A}_s \right] &=& 0 \; ;
\;\;\;{\rm here}, \; H_s=0, \;\;\; {\rm so \; that} \; 
[\hat{H}_s,\hat{\sigma}_z ]=0,
\label{}
\end{eqnarray}
so that
\begin{eqnarray}
\left[ \hat{H},\hat{A}_s \right] &=& 0 \; ;
\;\;\;{\rm here, indeed:} \; \left[\hat{H},\hat{\sigma}_z \right] = 0 .
\end{eqnarray}
\label{}
\end{subequations}
Just as in the previous case, the consequence is that starting with 
$|\psi_{spin}^{(0)} \rangle = |\sigma = 1\rangle$, say, the state will not get a component $|\sigma = -1\rangle$.

\section{Single measurements in Quantum Mechanics}
\label{single}

We first consider the measurement of an observable $\hat{A}$ using {\em one probe} and detecting a property of it. We are using the nomenclature introduced right after 
Eqs. (\ref{H(t),H0}).

For $\hat{A}$ we write the spectral representation
\begin{equation}
\hat{A} = \sum_n a_n {\mathbb P}_{a_n} \; ,
\label{sp. repr. A}
\end{equation}
where the eigenvalues $a_n$ are allowed to be degenerate and $\mathbb{P}_{a_n}$ are the eigenprojectors.

We assume the system to be coupled to a probe, considered, for simplicity, to be one-dimensional, whose position and momentum are represented by the Hermitean operators $\hat{Q}$ and $\hat{P}$.
The system-probe interaction is taken to be
\cite{Neumann-MathFounQuanMech:55}
\begin{equation}
\hat{V}(t) = \epsilon \; g(t) \hat{A} \hat{P}\; , \hspace{5mm}  t_1 > 0\; ,
\label{V_single}
\end{equation}
with an {\em arbitrary} interaction strength  \cite{Peres-QuanLimiDeteWeak:89} $\epsilon$. 

This interaction could be translated to the one for the Stern-Gerlach experiment discussed in Sec. \ref{pf(pz,t)} using the correspondence
$\hat{A} \Rightarrow \hat{\sigma}_z$, 
$\hat{P} \Rightarrow \hat{z}$,
$\hat{Q} \Rightarrow -\hat{p}_z$,
as illustrated in Fig. \ref{fig5} below. 
The delta function interaction of the previous section was generalized to
$g(t)$ (see Eq. (\ref{g(t)})), a narrow function with finite support, centered at $t=t_1$, so that 
\begin{subequations}
\begin{eqnarray}
\int_{0}^{t}g(t') dt' 
&\equiv& G(t), \\
G(0)&=&0, \;\;\;\;G(\infty)= 1 .
\end{eqnarray}
\label{g(t) 1}
\end{subequations}

We disregard the intrinsic evolution of the system and the probe, and assume that $\hat{V}(t)$ of Eq. (\ref{V_single}) represents the full Hamiltonian, i.e.,
\begin{equation}
\hat{H}(t) = \epsilon g(t) \hat{A}\hat{P}\; , \hspace{5mm}  t_1 > 0\; .
\label{H(t)_single}
\end{equation}

The evolution operator is then given by
\begin{equation}
\hat{U}(t)
= {\rm e}^{-\frac{i}{\hbar} \int_0^t \hat{H}(t')dt'}
= {\rm e}^{-\frac{i}{\hbar} \epsilon \; G(t)\hat{A}\hat{P}} .
\label{U_single}
\end{equation}
If the density operator of the system plus the probe at $t=0$ is the direct product
$
\rho^{(0)} = \rho_{s}^{(0)} \otimes \rho_{\pi}^{(0)}
$
($\pi$ stands for ``probe"), after the interaction has ceased to act, i.e.,
for $t \gg t_1$, it is given by
\begin{equation}
\rho^{(\hat{A})}_f
= \sum_{n  n' }
\mathbb{P}_{a_n}
\rho_{s}^{(0)}
\mathbb{P}_{a_{n'}}
(e^{-\frac{i}{\hbar}\epsilon a_n \hat{P}}
\rho_{\pi}^{(0)}
e^{\frac{i}{\hbar}\epsilon a_{n'} \hat{P}})
\; .
\label{rho_t>t1 single}
\end{equation}
From this expression we notice that, because of the interaction, the system and the probe are now correlated. 
Also notice the presence of the displacement operator
${\rm exp}(-(i/\hbar)\epsilon a_{n} \hat{P})$ in this last equation.

Now the idea is that at time $t>t_1$, i.e., after the system-probe interaction is over, we detect the probe position $\hat{Q}$ to obtain information on the system proper.
This we study in what follows.

\subsection{The $\hat{Q}$ probability density after the interaction}
\label{pf(Q) 0}

According to Born's rule, the $Q$ probability density for $t>t_1$ is given by
\begin{equation}
p_f^{(\hat{A})}(Q)
=
Tr \left(\rho_f^{(\hat{A})} \mathbb{P}_Q \right)
={\sum}_n
W_{a_n}^{(\hat{A})} \; p_0(Q-\epsilon a_n)  \; ,
\label{pf(Q)}
\end{equation}
where
\begin{equation}
W_{a_n}^{(\hat{A})}
={\rm Tr} ( \rho_s^{(0)} \mathbb{P}_{a_{n}} )
\label{Wan 0}
\end{equation}
is the {\em Born probability} for the result $a_n$ in the original system state, and
\begin{equation}
p_0(Q-\epsilon a_n)
=\langle Q-\epsilon a_n  |\rho_{\pi}^{(0)}| Q-\epsilon a_n  \rangle
\label{p0(Q-ean)}
\end{equation}
is the original $Q$ probability density $p_0(Q)$ (which has a width $= \sigma_Q$), but displaced by $\epsilon a_n$.

In this problem we may take the point of view that knowing the system state $\rho_s^{(0)}$ before the process, and thus $W_{a_n}^{(\hat{A})}$, 
we can predict the detectable quantity $p_f^{(\hat{A})}(Q)$. 
We may also adopt the more interesting viewpoint that, 
{\em detecting $p_f^{(\hat{A})}(Q)$, we can retrieve information on the system state}.
We examine this latter attitude below.

\begin{figure}[h]
\includegraphics[height=.3\textheight]{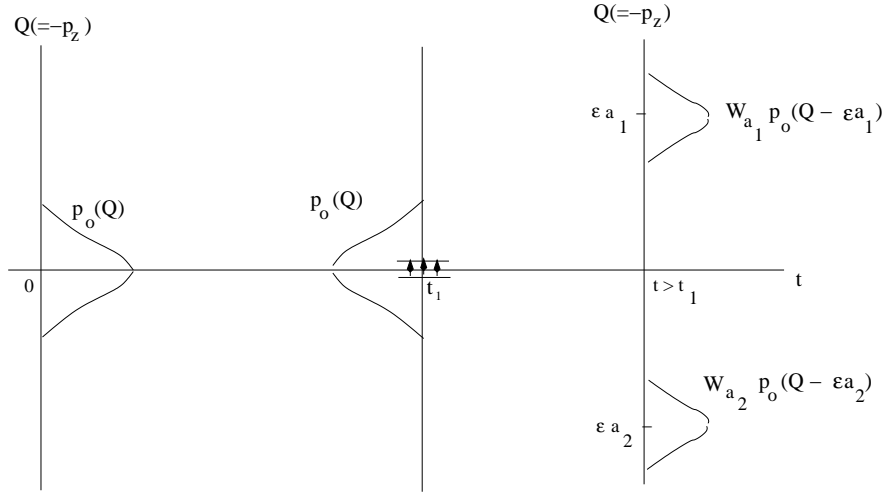}
\caption{The probability density $p_f^{(\hat{A})}(Q)$ of the probe degree of freedom $Q=-p_z$ as a function of time in the Stern-Gerlach experiment, as explained in the text.
This is basically Fig. \ref{fig4}, with the translation of variables 
$\hat{A} \Rightarrow \hat{\sigma}_z$, 
$\hat{P} \Rightarrow \hat{z}$,
$\hat{Q} \Rightarrow -\hat{p}_z$.}
\label{fig5}
\end{figure}

Before doing that, we illustrate in Fig. \ref{fig5} the result (\ref{pf(Q)}) for the case of the Stern-Gerlach experiment studied in the last section, by means of the translation given right before Eq. (\ref{g(t) 1}).

An illustrative example of a more general case is presented in Fig. \ref{fig6a,b}; 
for the values of the parameters indicated in the figure, Fig. 6a corresponds to the case of ``strong coupling", while Fig. 6b to that of ``weak coupling".
\begin{figure}[h]
  \includegraphics[height=.2\textheight]{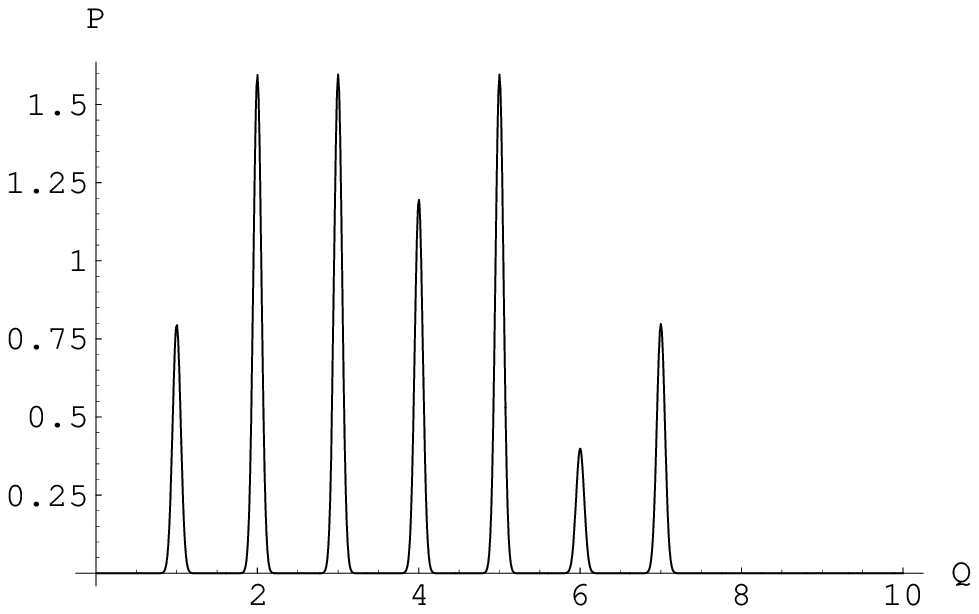}(a)
  \includegraphics[height=.2\textheight]{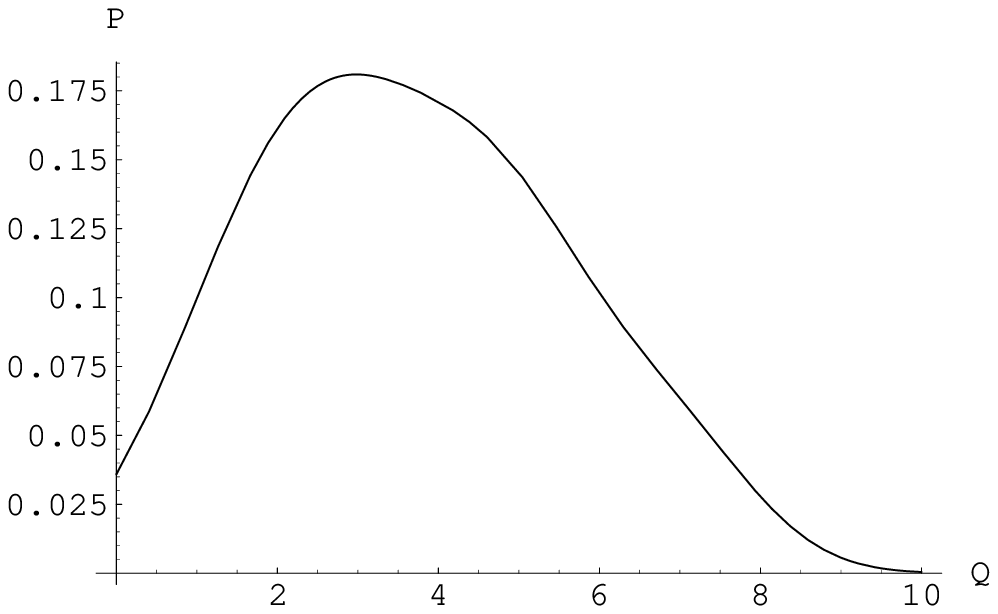}(b)
  \caption{Illustrative example of the probability density of the pointer position $Q$ of Eq. (\ref{pf(Q)}). 
We have assumed seven eigenvalues $a_n$ for the system observable $\hat{A}$, with Born probabilities 
[Eq. (\ref{Wan 0})] given by: 0.1, 0.2, 0.2, 0.15, 0.2, 0.05 and 0.1. We chose the particular values: 
(a) $\epsilon =1$ for the interaction strength and $\sigma_Q=0.05$ for the width of the probe state prior to the measurement; this is a case of ``strong coupling";
(b) $\epsilon =1$ for the interaction strength and $\sigma_Q=1$ for the width of the probe state prior to the measurement; 
this is a case of ``weak coupling".}
\label{fig6a,b}
\end{figure}

In the above scheme, what we detect is the probe-position probability 
$p_f^{(\hat{A})}(Q)$. As it can be seen from Fig. \ref{fig6a,b}, it is only in the idealized limit of very {\em strong coupling}, 
$\epsilon/\sigma_Q \gg 1$, that $p_f^{(\hat{A})}(Q)$ ``mirrors" the eigenvalues $a_n$ of the observable which we want to have information about. 
In this limit, $p_f^{(\hat{A})}(Q)$ {\em integrated around $a_n$ gives 
Born's probability $W_{a_n}^{(\hat{A})}$ of $a_n$}.
In this high-resolution limit this is thus the information we can retrieve about the system proper, by detecting $Q$.
We shall study below what information can be extracted from experiments with arbitrary resolution.
Surprisingly, we shall find cases where it is advantageous to use low resolution
(see Sec. \ref{state reconstruction}, last paragraph)!

\subsection{The average of $Q$ after the interaction}
\label{<Q> 0}

From Eqs. (\ref{pf(Q)}) and (\ref{Wan 0}) one finds \cite{aharonov_et_al} that the average of the probe position $\hat{Q}$ in units of $\epsilon$, after the interaction is over, is given by
\begin{equation}
\frac{1}{\epsilon}\langle \hat{Q} \rangle_f^{(\hat{A})}
= {\sum}_n  a_n W^{(\hat{A})}_{a_n}
= {\sum}_n  a_n {\rm Tr} (\rho_s^{(0)} \mathbb{P}_{a_n} )
= {\rm Tr} (\rho_s^{(0)} \hat{A} )
=\langle\hat{A}\rangle_0, \;\;\;\;\; \forall \epsilon \; .
\label{<Q>}
\end{equation}
We have assumed the original $Q$ distribution to be centered at $Q=0$.
As a result, detecting the average probe position $\langle \hat{Q}\rangle_f^{(\hat{A})}$ after the interaction is over {\em allows extracting 
the Born average} $\langle\hat{A}\rangle_0$ of the observable $\hat{A}$ in the original state of the system. 
It is remarkable that this result is valid for {\em arbitrary} coupling strength $\epsilon$.
For example, in the two situations illustrated in Fig. \ref{fig6a,b} we would obtain the same result for $\langle \hat{Q} \rangle_f^{(\hat{A})}/\epsilon$. 

Similar results can be found for higher-order moments.
E. g., for the second moment of $\hat{Q}$ one finds
\begin{equation}
\frac{1}{\epsilon^2}
\left[ \langle \hat{Q}^2 \rangle_f^{(\hat{A})} -\sigma_{Q}^2 \right] 
=  {\rm Tr} (\rho_s^{(0)} \hat{A}^2 ) 
= \langle \hat{A}^2 \rangle_0 \; , \;\;\; 
\forall \epsilon ,
\label{<Q^2>}
\end{equation}
implying that detecting $\langle \hat{Q}^2 \rangle_f^{(\hat{A})}$ and knowing 
$\sigma_{Q}^2$ allows extracting the second moment of the observable $\hat{A}$ in the original state of the system, i.e., $\langle \hat{A}^2 \rangle_0$.

More in general, if we detect the final $Q$ probability density (\ref{pf(Q)}) , i.e.,
\begin{equation}
p_f^{(\hat{A})}(Q) = \sum_n (\rho_s^{(0)})_{nn} \; p_0(Q-\epsilon a_n) \; ,
\label{pf(Q) vs rho_nn}
\end{equation}
we obtain information on the {\em diagonal elements} 
$(\rho_s^{(0)})_{nn}$ of the original density operator, but not on the off-diagonal ones.
Alternatively, we can write this result in terms of the characteristic function
\begin{equation}
\tilde{p}_f^{(\hat{A})}(k) = \left[\sum_n (\rho_s^{(0)})_{nn} 
{\rm e}^{ik\epsilon a_n}\right]
\tilde{p}_0(k) 
= \left\langle {\rm e}^{ik\epsilon \hat{A}}\right\rangle_0 \;
\tilde{p}_0(k) \; ,
\label{pf(k)}
\end{equation}
implying that if we detect $p_f^{(\hat{A})}(Q)$ and infer $\tilde{p}_f^{(\hat{A})}(k)$, we can extract $\left\langle {\rm e}^{ik\epsilon \hat{A}}\right\rangle_0$.
Results (\ref{<Q>}) and (\ref{<Q^2>}) are particular cases of Eq. (\ref{pf(k)}).

In Sec. \ref{state reconstruction} we shall find a procedure to extract all of the matrix elements of the original density operator, using the notion of successive measurements.

\subsection{Measuring projectors}
\label{measuring projectors}

A particular case of great interest is the measurement of a projector, like 
$\mathbb{P}_{a_{\nu}}$, so that the Hamiltonian of Eq. (\ref{H(t)_single}) becomes
\begin{equation}
\hat{H}(t) = \epsilon \; g(t) \hat{\mathbb{P}}_{a_{\nu}} \hat{P}\; , 
\hspace{5mm}  t_1 > 0\; .
\label{V_single 1}
\end{equation}
We designate the eigenvalues of $\hat{\mathbb{P}}_{a_{\nu}}$ by $\tau = 1,0$, and its eigenprojectors by
$(\hat{\mathbb{P}}_{a_{\nu}})_{\tau}$.
Then
\begin{subequations}
\begin{eqnarray}
(\hat{\mathbb{P}}_{a_{\nu}})_1 &=& \hat{\mathbb{P}}_{a_{\nu}} \; ;  \hspace{16mm}    
\hat{\mathbb{P}}_{a_{\nu}} (\hat{\mathbb{P}}_{a_{\nu}})_1 = 1 \cdot (\hat{\mathbb{P}}_{a_{\nu}})_1       \\
(\hat{\mathbb{P}}_{a_{\nu}})_0 &=& I - \hat{\mathbb{P}}_{a_{\nu}} \; ; \hspace{1cm} 
\hat{\mathbb{P}}_{a_{\nu}} (\hat{\mathbb{P}}_{a_{\nu}})_0 = 0 \cdot (\hat{\mathbb{P}}_{a_{\nu}})_0 \; .     
\end{eqnarray}
\label{Pa,nu}
\end{subequations}

\begin{figure}[h]
  \includegraphics[height=.15\textheight]{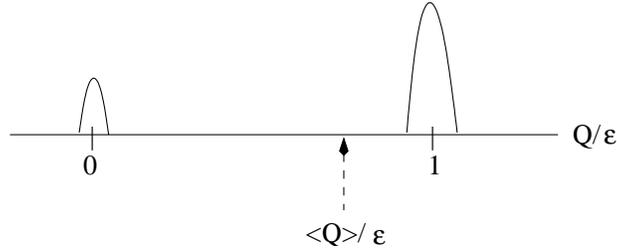}
  \caption{Illustrative example (qualitative) of the probability density of the probe position $Q$ of Eq. (\ref{pf(Q) for proj}) for the strong-coupling case.
The result consists of two peaks, centered at $Q/\epsilon =0$ and $Q/\epsilon =1$.
Also shown is the average of the probe position which, in units of $\epsilon$, is independent of $\epsilon$ and lies between 0 and 1.
}
\label{fig7}
\end{figure}

For these eigenvalues and eigenprojectors, the probe-position probability density of  Eq. (\ref{pf(Q)}) gives
\begin{equation}
p_f^{(\hat{\mathbb{P}}_{a_{\nu}})}(Q)
= {\rm Tr} \left[\rho_s^{(0)} (\hat{\mathbb{P}}_{a_{\nu}})_0 \right] \; p_0(Q)
+{\rm Tr} \left[\rho_s ^{(0)} (\hat{\mathbb{P}}_{a_{\nu}})_1 \right]  \; p_0(Q-\epsilon) \; .
\label{pf(Q) for proj}
\end{equation}
This result is illustrated in Fig. \ref{fig7} for the strong-coupling case, in which $p_f^{(\hat{\mathbb{P}}_{a_{\nu}})}(Q)$ consists of two peaks centered at $Q/\epsilon =0$ and $Q/\epsilon =1$.
From Eq. (\ref{pf(Q) for proj}), or from Eq. (\ref{<Q>}), we find 
\begin{equation}
\frac{1}{\epsilon}\langle \hat{Q} \rangle_f^{(\hat{\mathbb{P}}_{a_{\nu}})}
= \sum_{\tau =0}^1 \tau \; {\rm Tr}[\rho_s^{(0)}(\hat{\mathbb{P}}_{a_{\nu}})_{\tau}]
= \left\{
\begin{array}{l}
{\rm Tr} (\rho_s^{(0)} \hat{\mathbb{P}}_{a_{\nu}})
=W_{a_{\nu}}^{(\hat{A})} \\
{\rm Tr} [\rho_s^{(0)} (\hat{\mathbb{P}}_{a_{\nu}})_{\tau = 1}]
=W_{\tau=1}^{(\hat{\mathbb{P}}_{a_{\nu}})}
\end{array}
\right. \; ,
\label{<Q> for proj}
\end{equation}
a result independent of $\epsilon$. 
Thus the final average probe position, in units of $\epsilon$, gives {\em directly} the probability 
$W_{a_{\nu}}^{(\hat{A})}$ of $a_{\nu}$ of the observable $\hat{A}$ in the original state (see the first row of Eq. (\ref{<Q> for proj}), where we used 
$\sum_{\tau}\tau(\hat{\mathbb{P}}_{a_{\nu}})_{\tau}=\hat{\mathbb{P}}_{a_{\nu}}$); 
this is illustrated by the arrow in Fig. \ref{fig7}).
In contrast, we would not know how to extract the $W^{(\hat{A})}_{a_n}$'s from the sum in Eq. (\ref{<Q>}) if $\hat{A}$ is not a projector.
We can also say that inferring the
probability of the eigenvalue $\tau=1$ (second row in (\ref{<Q> for proj})), i.e., the probability of ``yes" of the observable $\hat{\mathbb{P}}_{a_{\nu}}$,
from  a detection of 
$\langle \hat{Q} \rangle_f^{(\hat{\mathbb{P}}_{a_{\nu}})}/ \epsilon$
(LHS of (\ref{<Q> for proj})),
is equivalent to retrieving the diagonal elements 
$\langle a_{\nu}|\rho_s^{(0)}|a_{\nu}\rangle$
of the original density operator of the system (first row in (\ref{<Q> for proj})).
As we shall see in Sec. \ref{state reconstruction}, the extension of this last idea to successive measurements will allow retrieving the full density operator.

\subsection{The reduced density operator of the system proper after the interaction}
\label{final reduced rho of system}

We compute the reduced density operator of the system proper after the interaction with the probe is over, by tracing $\rho_f^{(\hat{A})}$ of Eq. (\ref{rho_t>t1 single}) over the probe, with the result
\begin{subequations}
\begin{eqnarray}
\rho_{s,f}^{(\hat{A})} 
&=& \sum_{n n'} \left(\mathbb{P}_{a_n} \rho_s^{(0)} \mathbb{P}_{a_{n'}}\right) 
{\rm Tr}_{\pi}\left[ e^{-\frac{i}{\hbar}\epsilon a_n \hat{P}} \rho_{\pi}^{(0)} 
e^{\frac{i}{\hbar}\epsilon a_{n'} \hat{P}} \right] 
\label{reduced rho a}    \\
&=& \sum_{n,n'}
g(\epsilon(a_n - a_{n'}))\;
\mathbb{P}_{a_n}
\rho_{s}^{(0)}
\mathbb{P}_{a_{n'}} \; ,
\label{reduced rho b}
\end{eqnarray}
\label{reduced rho}
\end{subequations}
where we have defined the characteristic function of the probe momentum distribution as
\begin{equation}
g(\beta) 
= \left\langle {\rm e}^{-\frac{i}{\hbar}\beta \hat{P}}  \right\rangle_{\pi}^{(0)}
= {\rm Tr} \left[ \rho_{\pi}^{(0)} e^{-\frac{i}{\hbar}\beta \hat{P}} \right]; 
\; \; \;\;
\beta=\epsilon(a_n - a_{n'}) \; .
\label{g(beta)}
\end{equation}
As an example, for the particular case of a Gaussian state for the probe we have
\begin{subequations}
\begin{eqnarray}
\chi(Q)
&=& \frac{e^{-\frac{Q^2}{4\sigma^2_{Q}}}}{(2\pi\sigma^2_{Q})^{1/4}} 
\label{g(beta) gaussian a} \\
g(\epsilon(a_n - a_{n'}))
&=& \int \chi^{*}(Q-\epsilon a_{n'}) \chi(Q-\epsilon a_{n}) dQ  
\label{g(beta) gaussian b}\\
&=&e^{-\frac{\epsilon^2}{8\sigma_Q^2}(a_n - a_{n'})^2} 
= e^{-\frac12 \left(\frac{\epsilon \sigma_P}{\hbar}\right)^2
(a_n-a_{n'})^2} \; .
\label{g(beta) gaussian c}
\end{eqnarray}
\label{g(beta) gaussian}
\end{subequations}

The result (\ref{reduced rho}) is valid for an arbitrary state of the probe and an artbitrary coupling strength $\epsilon/\sigma_Q$.
In the strong-coupling limit $\epsilon/\sigma_Q \to \infty$, $\rho_{s,f}^{(\hat{A})}$ reduces to
\begin{equation}
\rho_{s,f}^{(\hat{A})}
= \sum_n \mathbb{P}_{a_n} \rho_s^{(0)} \; \mathbb{P}_{a_n}\; .
\label{reduced rho strong coupling}
\end{equation}
This is called the von Neumann-L\"uders rule, originally postulated by L\"uders \cite{Lueders-UEbeZust:51}, and then given a dynamical derivation in 
Ref. \cite{Bell+Nauenberg-MoraAspeQuanMech:66} using vNM.

We have thus reproduced the result of a {\em non-selective projective measurement}
\cite{Johansen-Quantheosuccproj:07} of the observable $\hat{A}$, as a limiting case of our general formalism.

Notice that $\rho_{s,f}^{(\hat{A})}$ and $\rho_s^{(0)}$ are not connected by a unitary transformation: indeed, their eigenvalues have changed.
For example, in the particular case in which the initial system state is the pure state
$\rho_s^{(0)}= |\psi_s^{(0)}\rangle \langle \psi_s^{(0)}|$, the initial 
eigenvalues are $1,0, \cdots,0$.
On the other hand, one eigenstate of $\rho_{s,f}^{(\hat{A})}$ is
$\mathbb{P}_{a_{\nu}}|\psi_s^{(0)}\rangle$, fulfilling the eigenvalue equation
\begin{equation}
\rho_{s,f}^{(\hat{A})}\left(\mathbb{P}_{a_{\nu}}|\psi_s^{(0)}\rangle\right)
= \langle \psi_s^{(0)}| \mathbb{P}_{a_{\nu}}| \psi_s^{(0)}\rangle
\left(\mathbb{P}_{a_{\nu}}|\psi_s^{(0)}\rangle\right) \; ,
\label{e-value of rho-f}
\end{equation}
so that the corresponding eigenvalue is 
$\langle \psi_s^{(0)} |\mathbb{P}_{a_{\nu}}| \psi_s^{(0)}\rangle$.
This is not a contradiction, because it is the density operator for the whole system, i.e., the system proper plus the probe, that evolves unitarily 
(see Eqn. (\ref{rho_t>t1 single})), while here we are dealing with the 
reduced density operator, which is the full density operator traced over the probe.

We computed above the reduced density matrix for the system proper, $\rho_{s,f}^{(\hat{A})}$, after the system-probe interaction;
then the probe is detected.
We now wish to make a model for the probe-detector ($\pi-D$) interaction,
taking place at some time $t_2>t_1$, and investigate if the resulting reduced density matrix for the system proper is still the same as the one given above, in Eq. (\ref{reduced rho}).
Assume that this new interaction {\em does not involve the system proper} $s$.
The final density operator after the interaction with the detector, to be called
$\rho_f^{(\rm{after \; inter. \; with \;  detector})}$, will contain a new evolution operator $U_{\pi D}$, that involves the 
probe and the detector, but not the system $s$.
Tracing $\rho_f^{(\rm{after \; inter. \; with \;  detector})}$
over the probe {\em and} the detector, we obtain
\begin{subequations}
\begin{eqnarray}
\rho_{s,f}^{(\rm{after \; inter. \; with \;  detector})}
&=& {\rm Tr}_{\pi D}\left(
\rho_f^{(\rm{after \; inter. \; with \;  detector})}
\right)\\
&=& \sum_{n n'} \left(\mathbb{P}_{a_n} \rho_s^{(0)} \mathbb{P}_{a_{n'}}\right) 
{\rm Tr}_{\pi D}
\left[ U_{\pi D}
e^{-\frac{i}{\hbar}\epsilon a_n \hat{P}} \rho_{\pi}^{(0)} 
e^{\frac{i}{\hbar}\epsilon a_{n'} \hat{P}} \hat{\rho}_D^{(0)} 
 U^{\dagger}_{\pi D} \right] 
\nonumber \\ \\
&=& \sum_{nn'}
g(\epsilon(a_n - a_{n'}))\;
\mathbb{P}_{a_n}
\rho_{s}^{(0)}
\mathbb{P}_{a_{n'}} \; ,
\end{eqnarray}
\label{reduced rho-s tracing pi and D}
\end{subequations}
giving the same answer as before, Eq. (\ref{reduced rho}).
We stress that this result holds when the probe-detector interaction does not involve $s$.

\section{Successive Measurements in Quantum Mechanics}
\label{successive}

We now generalize the problem of the previous section to describe the measurement of two observables in succession:
$\hat{A}$, as defined in Eq. (\ref{sp. repr. A}), at time $t_1$, and then
\begin{equation}
\hat{B} = \sum_m b_m \mathbb{P}_{b_m} \; ,
\label{sp. repr. B}
\end{equation}
(the $b_m$'s may also be degenerate), at some later time $t_2$.
For this purpose, we assume that we employ two probes, which are the ones that we detect; their momentum and coordinate operators are $\hat{P}_i$, $\hat{Q}_i$, $i=1,2$.
The interaction of the system with the probes defines the Hamiltonian
\begin{equation}
\hat{H} (t) = \epsilon_1 g_1(t) \hat{A} \hat{P}_1
+ \epsilon_2 g_2(t) \hat{B} \hat{P}_2 \; .
\label{V 2meas}
\end{equation}
Again, we have disregarded the intrinsic Hamiltonians of the system and of the two probes.
The functions $g_1(t)$ and $g_2(t)$ are narrow non-overlapping functions, centered around $t=t_1$ and $t=t_2$, respectively (see Eqs. (\ref{g(t) 1})), with $0 < t_1 < t_2 $.

The unitary evolution operator is given by
\begin{equation}
\hat{U}(t)
={\rm e}^{-\frac{i}{\hbar}\epsilon_2 G_2(t) \hat{B} \hat{P}_2}
{\rm e}^{-\frac{i}{\hbar}\epsilon_1 G_1(t) \hat{A} \hat{P}_1}.
\label{U 2meas}
\end{equation}

If the density operator of the system plus the probes at $t=0$ is assumed to be the direct product
$
\rho^{(0)} = \rho_{s} \otimes \rho_{\pi_1} \otimes \rho_{\pi_2}
$,
for $t \gg t_2$, i.e., after the second interaction has ceased to act, it is given by
\begin{eqnarray}
&&
\rho_f^{(\hat{B} \leftarrow \hat{A})}= \sum_{nn'mm'}
(\mathbb{P}_{b_{m}} \mathbb{P}_{a_{n}}  \rho_{s}^{(0)} \;\mathbb{P}_{a_{n'}} 
\mathbb{P}_{b_{m'}})
\nonumber \\
&& 
\hspace{5mm}\cdot \left(
e^{-\frac{i}{\hbar}\epsilon_1 a_n \hat{P_1}} \rho_{\pi_1}^{(0)}
e^{\frac{i}{\hbar}\epsilon_1 a_{n'}\hat{P_1}} \right)
\left(
e^{-\frac{i}{\hbar}\epsilon_2 b_m \hat{P_2}}  \rho_{\pi_2}^{(0)}
e^{\frac{i}{\hbar}\epsilon_2 b_{m'} \hat{P_2}}
\right) \; .
\label{rhof two probes}
\end{eqnarray}

At $t \gg t_2$ we detect the two probe positions and momenta in order to obtain information about the system.
Two examples are considered below.

We first detect the two probe positions $\hat{Q}_1$ and $\hat{Q}_2$. 
Their correlation can be calculated as
\begin{eqnarray}
\langle \hat{Q}_1 \hat{Q}_2 \rangle_f^{(\hat{B} \leftarrow \hat{A})}
= {\rm Tr} 
\left[\rho_f^{(\hat{B} \leftarrow \hat{A})}\hat{Q}_1 \hat{Q}_2 \right]\; ,
\label{<Q1Q2> 0}
\end{eqnarray}
with the result \cite{johansen-mello:2008,amir-pier}
\begin{equation}
\frac{\langle \hat{Q}_1 \hat{Q}_2 \rangle_f
^{(\hat{B} \leftarrow \hat{A})}}{\epsilon_1 \epsilon_2}
= \Re  \sum_{nm} a_{n} b_{m}
W^{(\hat{B} \leftarrow \hat{A})}_{b_{m}a_{n}}(\epsilon_1)\; ,
\label{<Q1Q2> 1}
\end{equation}
where $\Re$ stands for the real part.
We have defined
\begin{equation}
W^{(\hat{B} \leftarrow \hat{A})}_{b_{m}a_{n}}(\epsilon_1)
=\sum_{n'}
\lambda(\epsilon_1(a_n-a_{n'}))
{\rm Tr}\left[\rho_s^{(0)} (\mathbb{P}_{a_{n'}} \mathbb{P}_{b_{m}} 
\mathbb{P}_{a_{n}})\right]
\label{Wmn}
\end{equation}
and
\begin{subequations}
\begin{eqnarray}
\hspace{1cm}\lambda(\beta)
&=& g(\beta)+ 2 h(\beta) \; ,
\label{lambda(beta)}   \\
g(\beta) 
&=& \left\langle {\rm e}^{-\frac{i}{\hbar}\beta \hat{P}_1}\right\rangle_{\pi_1}^{(0)}, 
\label{g(beta)}  \\
h(\beta) 
&=&
\frac{1}{\beta}\left\langle  
{\rm e}^{-\frac{i}{2\hbar}\beta \hat{P}_1}  
\hat{Q}_1
{\rm e}^{-\frac{i}{2\hbar}\beta \hat{P}_1}
\right\rangle_{\pi_1}^{(0)} \; .
  \label{h(beta)}
\end{eqnarray}
\label{lambda,g,h}
\end{subequations}
Here, $\langle \cdots \rangle_{\pi_1}^{(0)}$ indicates an average over the initial state of probe 1.
Notice that 
$\langle \hat{Q}_1 \hat{Q}_2 \rangle_f^{(\hat{B} \leftarrow \hat{A})}$ 
may be a complicated function of $\epsilon_1$;
however, it is linear in $\epsilon_2$, the strength associated with the last measurement, just as for a single measurement we found, in Eq. (\ref{<Q>}), that
$\langle \hat{Q} \rangle_f^{(\hat{A})} \propto \epsilon$.
We also have the result (see Ref. \cite{amir-pier} for the relevant conditions)
\begin{equation}
\lambda(0)=1 \; .
\label{l(0)=1}
\end{equation}

The following comments are in order at this point.
Knowing the original system state $\rho_s^{(0)}$, the {\em auxiliary function} 
$\Re W^{(\hat{B} \leftarrow \hat{A})}_{b_{m}a_{n}}(\epsilon_1)$
appearing in Eq. (\ref{<Q1Q2> 1}) allows predicting the {\em detectable} quantity
$\langle \hat{Q}_1 \hat{Q}_2 \rangle_f^{(\hat{B} \leftarrow \hat{A})}$. 
It extends to two measurements the Born probability 
$W^{(\hat{A})}_{a_{n}}={\rm Tr} (\rho_s^{(0)} \mathbb{P}_{a_n})$ 
of Eq. (\ref{Wan 0}) which, for single measurements, allows predicting the detectable quantity 
$\langle \hat{Q} \rangle_f^{(\hat{A})}
=\epsilon {\sum}_n  a_n W^{(\hat{A})}_{a_n}
$ of Eq. (\ref{<Q>}).
More interestingly, {\em detecting} 
$\langle \hat{Q}_1 \hat{Q}_2 \rangle_f^{(\hat{B} \leftarrow \hat{A})}$, 
we investigate {\em what information can we retrieve on the system state}.
This is the point of view that will be taken in the next section.

As the next example, we consider again the same Hamiltonian of Eq. (\ref{V 2meas}) but, after the second interaction has acted, i.e., for $t \gg t_2$, we detect,
on a second sub-ensemble,
the momentum $\hat{P}_1$ of the first probe instead of its position, and the position $\hat{Q}_2$ of the second probe. 
The resulting correlation between $\hat{P}_1$ and $\hat{Q}_2$ is 
\cite{johansen-mello:2008,amir-pier}
\begin{equation}
\frac{1}{\epsilon_1 \epsilon_2}\langle \hat{P}_1 \hat{Q}_2 \rangle
 ^{(\hat{B} \leftarrow \hat{A})}
 =\frac{1}{2\sigma^2_{Q_1}}
\Im \sum_{nm}  a_{n} b_{m}
\tilde{W}^{(\hat{B} \leftarrow \hat{A})}_{b_{m}a_{n}}(\epsilon_1) \; ,
\label{<P1Q2> 0}
\end{equation}
where
\begin{equation}
\tilde{W}^{(\hat{B} \leftarrow \hat{A})}_{b_{m}a_{n}}(\epsilon_1)
=\sum_{n'}
\tilde{\lambda}(\epsilon_1(a_n-a_{n'}))
{\rm Tr}\left[\rho_s^{(0)} (\mathbb{P}_{a_{n'}} \mathbb{P}_{b_{m}} 
\mathbb{P}_{a_{n}})\right] \; ,
\label{Wtilde}
\end{equation}
and
\begin{subequations}
\begin{eqnarray}
\tilde{\lambda}(\beta)
&=&\frac{\bar{\lambda}(\beta)}{\bar{\lambda}(0)}\;,
\label{tilde-lambda} \\
\bar{\lambda}(\beta) 
&=& \frac{1}{\beta}\frac{\partial g(\beta)}{\partial \beta} \; .
\label{bar-lambda}
\end{eqnarray}
\label{tilde-bar-lambda}
\end{subequations}
Just as in the first example, the {\em auxiliary function} 
$\Im \tilde{W}^{(\hat{B} \leftarrow \hat{A})}_{b_{m}a_{n}}(\epsilon_1)$
allows predicting the {\em detectable} quantity
$\langle \hat{P}_1 \hat{Q}_2 \rangle_f^{(\hat{B} \leftarrow \hat{A})}$.
Alternatively, detecting
$\langle \hat{P}_1 \hat{Q}_2 \rangle_f^{(\hat{B} \leftarrow \hat{A})}$, 
we shall investigate {\em what information can we retrieve on the system state}.

\begin{figure}[h]
  \includegraphics[height=.2\textheight]{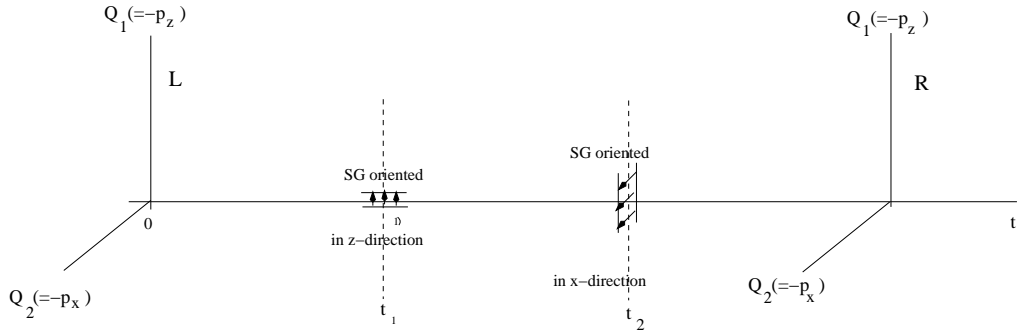}
  \caption{Illustration of the measurement of $\langle Q_1 Q_2 \rangle$ and 
$\langle P_1 Q_2 \rangle$ for a Stern-Gerlach arrangement. 
The correspondence between the various operators is given in 
Eqs. (\ref{translation to SG for 2 probes}).
Only the axes $Q_1$ and $Q_2$ are shown.
The procedure to measure $\langle Q_1 Q_2 \rangle$ would be:
a) send 1 atom from L to R; 
b) Measure $Q_1=-p_z$, $Q_2=-p_x$  $([\hat{Q}_1,\hat{Q}_2]=0)$ 
and construct $Q_1Q_2$;
c) Within an ensemble of $\cal{N}$ elements, construct 
$\langle Q_1Q_2\rangle = \frac{1}{\cal{N}}\sum_{i=1}^{\cal{N}}Q_1^{(i)}Q_2^{(i)}$.
The procedure to measure $\langle P_1 Q_2 \rangle$ would be
to construct 
$\langle P_1Q_2\rangle = \frac{1}{\cal{N}}\sum_{i=1}^{\cal{N}}P_1^{(i)}Q_2^{(i)}$
in another ensemble of $\cal{N}$ elements.}
\label{fig8}
\end{figure}

We illustrate in Fig. \ref{fig8} the measurements of 
Eqs. (\ref{<Q1Q2> 1}) and (\ref{<P1Q2> 0}) for the particular case of the Stern-Gerlach experiment studied in section \ref{stern-gerlach}, using the following translation of observables:
\begin{equation}
\begin{array}{cc}
\hat{A} \Rightarrow \hat{\sigma}_z & \hat{B} \Rightarrow \hat{\sigma}_x \\
\hat{P}_1 \Rightarrow \hat{z} & \hat{P}_2 \Rightarrow \hat{x} \\
\hat{Q}_1 \Rightarrow -\hat{p}_z & \hat{Q}_2 \Rightarrow -\hat{p}_x
\end{array}
\label{translation to SG for 2 probes}
\end{equation}

In what follows we examine some properties of the auxiliary functions
$W^{(\hat{B} \leftarrow \hat{A})}_{b_{m}a_{n}}(\epsilon_1)$
and
$
\tilde{W}^{(\hat{B} \leftarrow \hat{A})}_{b_{m}a_{n}}(\epsilon_1)$
defined in Eqs. (\ref{Wmn}) and (\ref{Wtilde}) 
(for more details, see Refs. \cite{johansen-mello:2008,amir-pier}).

1) In the {\em strong-coupling limit}, $\epsilon_1 \to \infty$, we find
\begin{equation}
\left.
\begin{array}{c}
W^{(\hat{B} \leftarrow \hat{A})}_{b_{m}a_{n}}(\epsilon_1) \\
\tilde{W}^{(\hat{B} \leftarrow \hat{A})}_{b_{m}a_{n}}(\epsilon_1)
\end{array}
\right\}
\to \;
{\cal W}_{b_{m}a_{n}}^{(\hat{B} \leftarrow \hat{A})}
\equiv {\rm Tr}\left[\rho_s^{(0)} (\mathbb{P}_{a_{n}} \mathbb{P} _{b_{m}}
\mathbb{P}_{a_{n}} )\right] \; .
\label{W e to infty}
\end{equation}
This is the joint probability distribution given by the so-called {\em Wigner's rule} 
\cite{Wigner-ProbMeas:63} (it is real and non-negative; notice its dependence on the order in which the two successive measurements are performed), which is obtained 
for {\em projective measurements} as 
\begin{eqnarray}
P(b_m, a_n) &=& P(b_m|a_n) P(a_n) 
=|\langle b_m|a_n \rangle|^2   |\langle a_n|\psi \rangle|^2 
\nonumber \\
&=& \langle \psi| \mathbb{P}_{a_{n}} \mathbb{P} _{b_{m}}
\mathbb{P}_{a_{n}} |\psi \rangle \; .
\label{wigner rule}
\end{eqnarray}
We have assumed a pure state and no degeneracy, and $P(b_m|a_n)$ denotes a conditional probability.

2) In the {\em weak-coupling limit}, $\epsilon_1 \to 0$, we find
\begin{equation}
\left.
\begin{array}{c}
W^{(\hat{B} \leftarrow \hat{A})}_{b_{m}a_{n}}(\epsilon_1) \\
\tilde{W}^{(\hat{B} \leftarrow \hat{A})}_{b_{m}a_{n}}(\epsilon_1)
\end{array}
\right\}
\to \;
{\cal K}^{(\hat{B} \leftarrow \hat{A})}_{b_{m}a_{n}} 
\equiv{\rm Tr}\left[\rho_s^{(0)} ( \mathbb{P}_{b_{m}} \mathbb{P}_{a_{n}})\right] .
\label{W e to 0}
\end{equation}
This is the Kirkwood's-Dirac joint {\em quasi-probability}, which is
complex, in general \cite{Kirkwood-QuanStatAlmoClas:33,Dirac-AnalBetwClasQuan:45,Steinberg-Condprobquantheo:95,Johansen+Luis-NoncWeakMeas:04}
(see also Ref. \cite{feynmann87}).

In this limit $\epsilon_1 \to 0$, the probes correlation can be written in various forms as follows
\begin{subequations}
\begin{eqnarray}
\frac{1}{\epsilon_1 \epsilon_2}\langle \hat{Q}_1 \hat{Q}_2 \rangle^{(\hat{B}\leftarrow \hat{A})}
&=& \sum_{nm} a_n b_m 
\frac12{\rm Tr}\left[\rho_s^{(0)} 
(\mathbb{P}_{b_{m}} \mathbb{P}_{a_{n}} + \mathbb{P}_{a_{n}} \mathbb{P}_{b_{m}} ) \right] 
\label{<Q1Q2> e1 to 0 a}  \\
&=&  \sum_{nm} a_n b_m  
W^{MH}_{b_{m} a_{n}} 
= \sum_{nm} a_n b_m  
\langle \hat{S}_{mn} \rangle_0
\label{<Q1Q2> e1 to 0 b}  \\
&=&  \frac12 {\rm Tr}\left[\rho_s^{(0)} (\hat{B}\hat{A}+\hat{A}\hat{B} )\right]
\label{<Q1Q2> e1 to 0 c}
\end{eqnarray}
where
\begin{eqnarray}
W^{MH}_{b_{m} a_{n}}
&=& \frac12{\rm Tr}\left[\rho_s^{(0)} 
(\mathbb{P}_{b_{m}} \mathbb{P}_{a_{n}} +\mathbb {P}_{a_{n}} \mathbb{P}_{b_{m}})
\right] 
= \langle \hat{S}_{mn} \rangle_0 \; ,
\label{W(MH)} \\
\hat{S}_{mn} &\equiv&
\frac12
\left(
\mathbb{P}_{b_{m}} \mathbb{P}_{a_{n}}
+\mathbb{P}_{a_{n}}\mathbb{P}_{b_{m}} 
\right) \; ;
\label{Smn} 
\end{eqnarray}
\label{<Q1Q2> e1 to 0}
\end{subequations}
$W^{MH}_{b_{m} a_{n}}$ is the real part of the Kirkwood quasi-probability distribution
\cite{Kirkwood-QuanStatAlmoClas:33,Dirac-AnalBetwClasQuan:45,Steinberg-Condprobquantheo:95,Johansen+Luis-NoncWeakMeas:04}, also called the 
"Margenau-Hill (MH) distribution" \cite{Margenau+Hill-CorrbetwMeasQuan:61}: 
it may take negative values, and thus cannot be regarded as a joint probability in the classical sense.

We remark that, if $[\mathbb{P}_{b_{m}}, \mathbb{P}_{a_{n}}] \neq 0$, then the operator $\hat{S}_{mn}$ of Eq. (\ref{Smn}) has at least one negative e-value.
For this pair of variables, i.e., $a_n$, $b_m$, and for the particular
state of the system which is the eigenstate that gives rise to this negative eigenvalue, the Margenau-Hill distribution of Eq. (\ref{W(MH)}) is negative, i.e.,
$W^{MH}_{b_{m} a_{n}}=\langle \hat{S}_{mn} \rangle_0<0$, 
and the probes correlation
$\langle \hat{Q}_1 \hat{Q}_2 \rangle/\epsilon_1\epsilon_2$ 
{\em may} lie outside the  range 
$[(a_{n}b_{m})_{min}, (a_{n}b_{m})_{max}]$ (see Eq. (\ref{<Q1Q2> e1 to 0 b})).
We illustrate this point with an example.

Consider a Hilbert space of dimensionality $N=2$, and two operators $\hat{A}$ and 
$\hat{B}$ having the following eigenvalues and eigenvectors:
\begin{subequations}
\begin{eqnarray}
a_n &=& 1,0 \; ; \hspace{1cm}
| 1 \rangle = 
\left[
\begin{array}{c}
1 \\
0
\end{array} 
\right], \hspace{9mm}
| 0 \rangle = 
\left[
\begin{array}{c}
0 \\
1
\end{array} 
\right],
\\
b_m &=& 1,0  \; ; \hspace{7mm} \;\;\;
| 1 ) = 
\frac{1}{\sqrt{2}}
\left[
\begin{array}{c}
1 \\
1
\end{array} 
\right], \;\;
| 0 ) = 
\frac{1}{\sqrt{2}}
\left[
\begin{array}{r}
1 \\
-1
\end{array} 
\right]. 
\end{eqnarray}
\label{e-values,e-vectors N2 for MH<0}
\end{subequations}
From Eq. (\ref{<Q1Q2> e1 to 0 b}) we have
\begin{equation}
\frac{1}{\epsilon_1 \epsilon_2}
\langle \hat{Q}_1 \hat{Q}_2 \rangle^{(\hat{B}\leftarrow \hat{A})}
= \sum_{n,m=0}^1 a_n b_m \langle S_{mn} \rangle_0 = \langle S_{11} \rangle_0 \; .
\label{<Q1Q2> e=0 N=2}
\end{equation}
For the two bases of Eq. (\ref{e-values,e-vectors N2 for MH<0}), we find 
$S_{11}$ and its eigenvalues as
\begin{equation}
S_{11}
= \frac14
\left[
\begin{array}{cc}
2 & 1 \\
1 & 0
\end{array}
\right]
\;\;\; 
\begin{array}{c}
\lambda_+ = \frac14 (1+ \sqrt{2}) > 0 \Rightarrow | \psi_+ \rangle \\
\lambda_- = \frac14 (1- \sqrt{2}) < 0  \Rightarrow | \psi_{-} \rangle
\end{array}
\label{S11 and its eigenvalues}
\end{equation}
For the state $| \psi_{-}\rangle$, the position-position correlation of 
Eq. (\ref{<Q1Q2> e=0 N=2}) becomes
\begin{equation}
\frac{1}{\epsilon_1 \epsilon_2}\langle \hat{Q}_1 \hat{Q}_2 \rangle^{(\hat{B}\leftarrow \hat{A})} 
= \left\langle \psi_- \right|S_{11}\left| \psi_-  \right\rangle
=\frac14 (1- \sqrt{2}) < 0
\label{}
\end{equation}
which lies outside the interval defined by the
possible values $0,1$ of the product $ss'$.

3) For an intermediate, {\em arbitrary} $\epsilon_1$, 
$W^{(\hat{B} \leftarrow \hat{A})}_{b_{m}a_{n}}(\epsilon_1)$ and 
$\tilde{W}^{(\hat{B} \leftarrow \hat{A})}_{b_{m}a_{n}}(\epsilon_1)$
can be regarded, as already noted above, as two auxiliary functions, 
$\Re W^{(\hat{B} \leftarrow \hat{A})}_{b_{m}a_{n}}(\epsilon_1)$
being taylored for predicting
$\langle \hat{Q}_1 \hat{Q}_2 \rangle$, and 
$\Im \tilde{W}^{(\hat{B} \leftarrow \hat{A})}_{b_{m}a_{n}}(\epsilon_1)$ for 
$\langle \hat{P}_1 \hat{Q}_2 \rangle$.
In the next section we shall see that for some special $\hat{A}$'s and $\hat{B}$'s  
we can realize the inverse case:
from the measurable quantities 
$\langle \hat{Q}_1 \hat{Q}_2 \rangle$ and 
$\langle \hat{P}_1 \hat{Q}_2 \rangle$
we can reconstruct $\rho_s^{(0)}$.

4) If the projectors $\mathbb{P}_{a_{n}}$, $\mathbb{P}_{b_{m}}$ appearing in 
Eqs. (\ref{Wmn}) and (\ref{Wtilde}) commute, i.e., 
$
\left[\mathbb{P}_{a_{n}}, \mathbb{P}_{b_{m}}\right] = 0, \;\;\forall n,m\; ,
$
then we find, for {\em arbitrary} $\epsilon_1$
\begin{equation}
W^{(\hat{B} \leftarrow \hat{A})}_{b_{m}a_{n}}(\epsilon_1)
=\tilde{W}^{(\hat{B} \leftarrow \hat{A})}_{b_{m}a_{n}}(\epsilon_1)
={\rm Tr}\left[\rho_s^{(0)} ( \mathbb{P}_{b_{m}}\mathbb{P}_{a_{n}} )\right], 
\;\;\;\; \forall \epsilon_1 \; .
\label{W=tilde W in comm case}
\end{equation}
This result is the standard, real and non-negative, quantum-mechanical definition of the joint probability of $a_n$ and $b_m$ for commuting observables.

We also find that the correlation of the two probe positions measured in units of 
$\epsilon_1 \epsilon_2$ coincides, for an arbitrary coupling strength $\epsilon_1$, with the standard result for the correlation of the two observables $\hat{A}$ and $\hat{B}$, i.e., 
\begin{equation}
\frac{1}{\epsilon_1 \epsilon_2 }\langle Q_1 Q_2 \rangle
=  {\rm Tr}\left[\rho_s^{(0)} (\hat{A}\hat{B})\right]\;, \;\;\; \forall \epsilon_1 \; .
\label{<Q1Q2>=<AB> in comm case}
\end{equation}

5) For the particular case in which  $\pi_1$ is described by a pure Gaussian state,
we find (see also Eqs. (\ref{g(beta) gaussian})) 
\begin{subequations}
\begin{eqnarray}
\lambda(\beta) &=& \tilde{\lambda}(\beta) 
=g(\beta)
=e^{-\frac{\beta^2}{8\sigma_{Q_1}^2}}
\label{} \\
W^{(\hat{B} \leftarrow \hat{A})}_{b_{m}a_{n}}(\epsilon_1)
&=&\tilde{W}^{(\hat{B} \leftarrow \hat{A})}_{b_{m}a_{n}}(\epsilon_1)\; .
\end{eqnarray}
\label{g,gtilde,gaussian case}
\end{subequations}
This is the case studied in Ref. \cite{johansen-mello:2008}.

6) As a particular situation of property 4) above, suppose we 
{\em measure successively the same observable} $\hat{A}$. 
We thus set $\hat{B}=\hat{A}$ in the formalism.
From Eq. (\ref{Wmn}) we find
\begin{eqnarray}
W^{(\hat{A} \leftarrow \hat{A})}_{a_{\bar{n}}a_{n}}(\epsilon_1)
&=&\sum_{n'}
\lambda(\epsilon_1(a_n-a_{n'}))
{\rm Tr}_s\left[\rho_s^{(0)} (\mathbb{P}_{a_{n'}} \mathbb{P}_{a_{\bar{n}}} 
\mathbb{P}_{a_{n}})\right] 
\nonumber \\
&=& 
{\rm Tr}_s \left(\rho_s^{(0)}\mathbb{P}_{a_{n}}\right)
\delta_{a_{\bar{n}},a_n} \; .
\label{W for B=A}
\end{eqnarray}
Eq. (\ref{<Q1Q2> 1}) then gives
\begin{equation}
\frac{\langle Q_1 Q_2 \rangle_f^{(\hat{A} \leftarrow \hat{A})}}
{\epsilon_1 \epsilon_2 }
= \sum_{n,\bar{n}} a_n a_{\bar{n}} \Re W^{(\hat{A} \leftarrow \hat{A})}_{a_{\bar{n}}a_{n}}(\epsilon_1)
= \sum_n W_{a_n}^{(\hat{A})} a_n^2 
= \langle \hat{A^2}\rangle_0 \; .
\label{<Q1Q2> for B=A}
\end{equation}

For simplicity, we restrict ourselves to the case in which, at $t=0$, the system proper $s$ and the two probes $\pi_1$, $\pi_2$ are described by pure states, and
\begin{equation}
|\Psi \rangle_0 =  |\psi \rangle_s^{(0)}  |\chi \rangle_{\pi_1}^{(0)} 
|\chi \rangle_{\pi_2}^{(0)} \; .
\label{Psi0}
\end{equation}
Then, for $t \gg t_2$, i.e., after the second interaction, the state vector is given by
\begin{subequations}
\begin{eqnarray}
|\Psi \rangle_f 
&=& {\rm e}^{-\frac{i}{\hbar}\epsilon_2  \hat{A} \hat{P}_2}
{\rm e}^{-\frac{i}{\hbar}\epsilon_1 \hat{A} \hat{P}_1}
|\Psi \rangle_{0}
\label{Psi_f a} \\
&=& \sum_n \left(\mathbb{P}_{a_n} |\psi \rangle_s^{(0)}\right)
\left({\rm e}^{-\frac{i}{\hbar}\epsilon_2  a_n \hat{P}_1} 
|\chi \rangle_{\pi_1}^{(0)}\right)
\left({\rm e}^{-\frac{i}{\hbar}\epsilon_1  a_n \hat{P}_2} 
|\chi \rangle_{\pi_2}^{(0)}\right) \; .
\label{Psi_f b}
\end{eqnarray}
\label{Psi_f}
\end{subequations}
The joint probability density (jpd) of the eigenvalues $Q_1, Q_2$ of the two position operators for times $t>t_2$, when the two interactions have ceased to act, is then
\begin{subequations}
\begin{eqnarray}
p_f(Q_1,Q_2)
&=&
_f\langle \Psi | 
\mathbb{P}_{Q_1} \mathbb{P}_{Q_2}
|\Psi  \rangle _f
\label{p(Q1,Q2) a} \\
&=& \sum_n
W_{a_n}^{(\hat{A})} \left|\chi_{\pi_1}^{(0)}(Q_1-\epsilon_1 a_n) \right|^2 
\left|\chi_{\pi_2}^{(0)}(Q_2-\epsilon_2 a_n) \right|^2 
\label{p(Q1,Q2) b} \\
&=& \sum_n
W_{a_n}^{(\hat{A})}
\frac{{\rm e}^{-\frac{(Q_1-\epsilon_1 a_n)^2}{2\sigma_{Q_1}^2}}}
{\sqrt{2\pi \sigma_{Q_1}^2}}
\frac{{\rm e}^{-\frac{(Q_2-\epsilon_2 a_n)^2}{2\sigma_{Q_2}^2}}}
{\sqrt{2\pi \sigma_{Q_2}^2}} \; .
\label{p(Q1,Q2) c} 
\end{eqnarray}
\label{p(Q1,Q2) 1}
\end{subequations}
In Eq. (\ref{p(Q1,Q2) c}) we have assumed the original pure states for the probes to be Gaussian.
Also,
\begin{equation}
W_{a_n}^{(\hat{A})}
=\; ^{(0)}{_s\langle} \psi |\hat{\mathbb P}_{a_n}|\psi\rangle_s^{(0)}
\label{Wan}
\end{equation}
is the Born probability for the value $a_n$ in the original system state, and we wrote
\begin{equation}
\langle Q_1 | {\rm e}^{-\frac{i}{\hbar}\epsilon_1 a_n \hat{P}_1}
 |\chi \rangle_{\pi_1}^{(0)}
= \chi_{\pi_1}^{(0)}(Q_1-\epsilon_1 a_n) \; ,
\label{chi(Q-ea)}
\end{equation}
and similarly for probe $\pi_2$.

Clearly, result (\ref{<Q1Q2> for B=A}) can be verified from the jpd of 
Eq. (\ref{p(Q1,Q2) c}).
From this jpd we also obtain
\begin{subequations}
\begin{eqnarray}
\langle \hat{Q}_i\rangle 
&=& \epsilon_i \langle \hat{A} \rangle_0 \; , \hspace{15mm} i=1,2,
\label{<Qi> for A=B} \\
\langle \hat{Q}_i^2\rangle 
&=& \epsilon_i^2 \langle \hat{A}^2 \rangle_0
+ \sigma_{Q_i}^2 \; , \;\;\; i=1,2.
\label{<(Qi)2> for A=B}
\end{eqnarray}
\label{<Qi>,<(Qi)2> for A=B}
\end{subequations}

\begin{figure}[h]
  \includegraphics[height=.3\textheight]{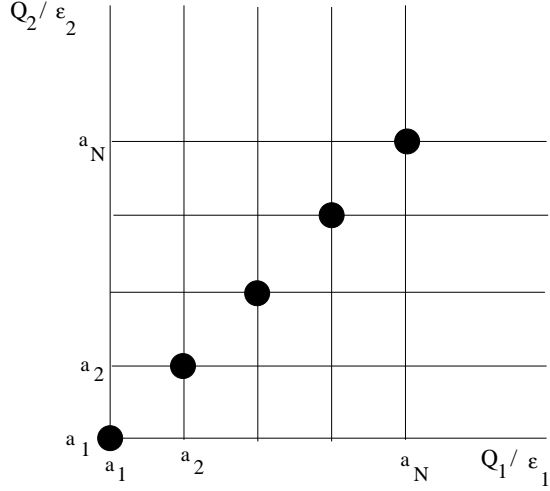}
  \caption{Schematic illustration of the jpd of the two probe positions $p_f(Q_1,Q_2)$ of Eq. (\ref{p(Q1,Q2) c}),
when we measure subsequently the {\em same} observable $\hat{A}$.  
The illustration is for the strong-coupling limit $\sigma_{Q_i}/\epsilon_i \ll 1$, 
in which $Q_1$ and $Q_2$ are strongly correlated.
}
\label{fig9}
\end{figure}

It is useful to consider the correlation coefficient between the two probe positions, which is defined as
\begin{subequations}
\begin{eqnarray}
C(Q_1,Q_2)
&=& \frac{\langle Q_1 Q_2 \rangle - \langle Q_1 \rangle \langle Q_2 \rangle}
{\sqrt{[\langle Q_1^2 \rangle - \langle Q_1 \rangle^2]
[\langle Q_2^2 \rangle - \langle Q_2 \rangle^2]}} 
\label{Q1Q2 corr. coeff. for B=A a} \\
&=& \frac{({\rm var}\hat{A})_0}
{\sqrt{({\rm var}\hat{A})_0 + \left(\frac{\sigma_{Q_1}}{\epsilon_1}\right)^2}
\sqrt{({\rm var}\hat{A})_0 + \left(\frac{\sigma_{Q_2}}{\epsilon_2}\right)^2}}
\label{Q1Q2 corr. coeff. for B=A b} \\
&\to& 1, \;\;\;\;\; \rm{as} \;\;\;\;\;\sigma_{Q_i} / \epsilon_i \to 0 \; .
\label{Q1Q2 corr. coeff. for B=A c}
\end{eqnarray}
\label{Q1Q2 corr. coeff. for B=A}
\end{subequations}
In the second line, (\ref{Q1Q2 corr. coeff. for B=A b}), we have used results (\ref{<Qi>,<(Qi)2> for A=B}) for a pure Gaussian state of the probes.
As a result, in the strong-coupling limit $\sigma_{Q_i}/\epsilon_i \to 0$,
the outcomes for the probe position 1 and probe position 2 become completely correlated, which is the result we would have expected.
This is illustrated schematically in Fig. \ref{fig9}.

7) Relation of successive measurements to {\em weak values}.

The weak value of the observable $\hat{A}$, with pre-selection $|\psi \rangle$ 
and post-selection $|\phi \rangle$, is defined as
\cite{aharonov_et_al,Johansen+Luis-NoncWeakMeas:04}
\begin{subequations}
\begin{eqnarray}
(\hat{A})_W 
&=& \frac{\langle \phi|\hat{A}|\psi \rangle}
{\langle \phi |\psi \rangle}
\label{WV A a}   \\
&=& 
\frac{\langle \psi | \phi \rangle \langle \phi|\hat{A}|\psi \rangle}
{\langle \psi | \phi \rangle\langle \phi |\psi \rangle}
=
\frac{\langle \psi | \hat{\mathbb{P}}_{\phi}\hat{A}|\psi \rangle}
{\langle \psi | \hat{\mathbb{P}}_{\phi} | \psi \rangle} 
\label{WV A b}   \\
&\Rightarrow& 
\frac{ {\rm Tr}\left(\hat{\rho}_s^{(0)}\hat{\mathbb{P}}_{\phi}\hat{A} \right)}
{{\rm Tr}\left(\hat{\rho}_s^{(0)}\hat{\mathbb{P}}_{\phi}\right) } 
\label{WV A c}   \\
&=& \sum_n a_n \frac{\langle \hat{\mathbb{P}}_{\phi} \hat{\mathbb{P}}_{a_n}\rangle}
{\langle \hat{\mathbb{P}}_{\phi} \rangle} 
\label{WV A d}
\end{eqnarray}
\label{WV A}
\end{subequations}
We see that the weak value can be regarded as the
correlation function between the observable $\hat{A}$
and the projector $\hat{\mathbb{P}}_{\phi}$ \cite{Johansen04,johansen-arxiv-09}.
Eq. (\ref{WV A c}) gives the generalization of the definition for a state
described by a density operator.
Eq. (\ref{WV A d}) expresses the weak value as the sum over the states of $a_n$
times a ``complex probability" of $a_n$ conditioned by $\phi$.

Consider now a successive measurement experiment in which the two observables are
$\hat{A}$ and $\hat{\mathbb{P}}_{\phi}$.
The Hamiltonian of Eq. (\ref{V 2meas}) becomes
\begin{equation}
\hat{H}(t) = \epsilon_1 g_1(t) \hat{A} \hat{P}_1
 + \epsilon_2 g_2(t) \hat{\mathbb{P}}_{\phi} \hat{P}_2, \;\;\;\;\;
0 < t_1 < t_2.
\label{H(t) for WV}
\end{equation}
One can show 
\cite{johansen-mello-arxiv-09,johansen_to_be_published}
that the position-position and momentum-position correlation of the two probes are related to the real and imaginary parts of the weak value as
\begin{subequations}
\begin{eqnarray}
\lim_{\epsilon_1 \to 0}
\frac{\langle \hat{Q}_1 \hat{Q}_2 \rangle^{(\hat{\mathbb{P}}_{\phi}\Leftarrow \hat{A})}}
{\epsilon_1\langle \hat{Q}_2\rangle^{(\hat{\mathbb{P}}_{\phi}\Leftarrow \hat{A})}}
&=& \frac12 
\frac{\langle \hat{A} \hat{\mathbb{P}}_{\phi} + \hat{\mathbb{P}}_{\phi}  \hat{A}\rangle}
{\langle \hat{\mathbb{P}}_{\phi} \rangle}
= \Re \left[(\hat{A})_W \right] \; ,
\label{<Q1Q2> and WV}   \\
\lim_{\epsilon_1 \to 0}
\frac{\langle \hat{P}_1 \hat{Q}_2 \rangle^{(\hat{\mathbb{P}}_{\phi}\Leftarrow \hat{A})}}
{\epsilon_1\langle \hat{Q}_2\rangle^{(\hat{\mathbb{P}}_{\phi}\Leftarrow \hat{A})}}
&=& \frac{1}{2\sigma^2_{Q_1}} 
\frac{\langle \hat{\mathbb{P}}_{\phi} \hat{A} - \hat{A} \hat{\mathbb{P}}_{\phi} \rangle}
{2 i \langle \hat{\mathbb{P}}_{\phi} \rangle}
= 2 \sigma^2_{P_1} \Im \left[(\hat{A})_W \right] \; .
\label{<P1Q2> and WV}
\end{eqnarray}
\label{succ meas and WV}
\end{subequations}

\section{STATE RECONSTRUCTION SCHEME BASED ON SUCCESSIVE MEASUREMENTS}
\label{state reconstruction}

We now use the above formalism to describe a state tomography scheme.
We build on previous work
\cite{Johansen-Quantheosuccproj:07,johansen-mello:2008}
to identify a set of observables which, when measured in succession, provide complete information about the state of a quantum system described in an $N$-dimensional Hilbert space.

For this purpose we consider, in our Hilbert space, two orthonormal bases, whose vectors are denoted by $|k \rangle$ and $|\mu)$, respectively, with 
$k, \mu=1,\ldots, N$.
Latin letters will be used to denote the first basis while Greek letters will be used for the second basis.
Given $k$, there is only one vector; given $\mu$, there is also only one vector: i.e., we have no degeneracy.

We assume the two bases are mutually non-orthogonal, i.e.,
\begin{equation}
\langle k |\mu ) \neq 0, \hspace{5mm} \forall k, \mu .
\label{mutual non-orthogonality}
\end{equation}
Then the two bases have no common eigenvectors.
\begin{figure}[h]
  \includegraphics[height=.2\textheight]{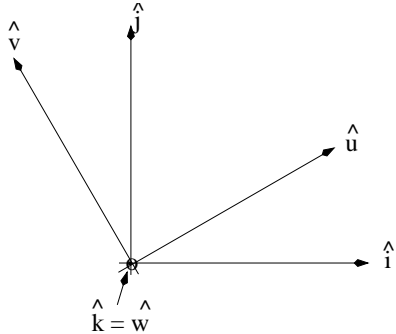}
  \caption{Illustration, for $N=2$, of the fact that if the two orthonormal bases have one common eigenvector, they cannot be mutually non-orthogonal.
}
\label{fig10}
\end{figure}
This is illustrated in Fig. \ref{fig10} for dimensionality $N=3$.
The basic vectors of the first basis are $\hat{i},\hat{j},\hat{k}$.
Those of the second basis are $\hat{u},\hat{v},\hat{w}$.
Assume they have one common eigenvector: e.g., $\hat{k}=\hat{w}$. 
Then $\hat{u},\hat{v} \perp \hat{k}$ and 
$\hat{i},\hat{j} \perp \hat{w}$,
contradicting the assumption that the two bases have no mutually orthogonal eigenvectors.

We now consider a successive-measurement experiment in which the two observables are the rank-one projectors $\mathbb{P}_{k}$ and $\mathbb{P}_{\mu}$ onto the $k$- and $\mu$-state of the first and second basis, respectively.
The Hamiltonian of Eq. (\ref{V 2meas}) becomes
\begin{equation}
\hat{H} (t) 
= \epsilon_1 g_1 (t) \mathbb{P}_{k} \hat{P}_1
+ \epsilon_2 g_2 (t) \mathbb{P}_{\mu} \hat{P}_2 \; .
\label{V(t)proj}
\end{equation}
The projectors $\mathbb{P}_{k}$ and $\mathbb{P}_{\mu}$ possess two eigenvalues: 
$0$ and $1$.
We denote by $\tau$ and $\sigma$ the eigenvalues of $\mathbb{P}_{k}$ and  $\mathbb{P}_{\mu}$, respectively, and the corresponding eigenprojectors by
$(\mathbb{P}_{k})_{\tau}$ and $(\mathbb{P}_{\mu})_{\sigma}$.
They satisfy relations of the type presented in Eqs. (\ref{Pa,nu}).

In the present case, Eq. (\ref{<Q1Q2> 1}) for the probes position-position correlation function gives
\begin{subequations}
\begin{eqnarray}
\frac{1}{\epsilon_1 \epsilon_2}
\langle \hat{Q}_1 \hat{Q}_2 \rangle
^{(
\mathbb{P}_{\mu} \leftarrow \mathbb{P}_{k}
)}
&=& \Re \sum_{\tau,\sigma = 0}^1 \tau \sigma \;
W^{( \mathbb{P}_{\mu} \leftarrow \mathbb{P}_{k})}
_{\sigma \tau} (\epsilon_1)
\label{<Q1Q2> P P a} 
\\
&=& 
\Re W^{( \mathbb{P}_{\mu} \leftarrow \mathbb{P}_{k})}_{1 1} (\epsilon_1).
\label{<Q1Q2> P P b}
\end{eqnarray}
\label{<Q1Q2> P P}
\end{subequations}
In Eq. (\ref{<Q1Q2> P P a}), 
$W^{( \mathbb{P}_{\mu} \leftarrow \mathbb{P}_{k})}
_{\sigma \tau} (\epsilon_1)$
is the particular case of the quantity
$W^{(\hat{B} \leftarrow \hat{A})}_{b_{m}a_{n}}(\epsilon_1)$
of Eq. (\ref{Wmn}) 
when $\hat{A}$, $\hat{B}$, $a_n$, and $b_m$ are replaced by $\mathbb{P}_{k}$, $\mathbb{P}_{\mu}$, $\tau$, and $\sigma$, respectively, i.e.,
\begin{equation}
W^{(\mathbb{P}_{\mu} \leftarrow \mathbb{P}_{k})}
_{\sigma \tau} (\epsilon_1)
=\sum_{\tau'=0}^1 \lambda(\epsilon_1(\tau-\tau'))
{\rm Tr}\left[\rho_s  (\mathbb{P}_{k})_{\tau'} (\mathbb{P}_{\mu})_{\sigma} 
(\mathbb{P}_{k})_{\tau} \right] \; ,
\end{equation}
and, in particular,
\begin{subequations}
\begin{eqnarray}
W^{( \mathbb{P}_{\mu} \leftarrow \mathbb{P}_{k})}_{1 1} (\epsilon_1)
&=&
{\rm Tr}(\rho_{s}^{(0)}\mathbb{P}_{k}\mathbb{P}_{\mu} \mathbb{P}_{k})
+\lambda(\epsilon_1) 
\sum_{k' (\neq k)} {\rm Tr} (\rho_{s}^{(0)}
\mathbb{P}_{k'}\mathbb{P}_{\mu} \mathbb{P}_{k}) \; ,
\label{W11 1 a} \\
&=& \sum_{k'}G_{kk'}(\epsilon_1) {\rm Tr} (\rho_{s}^{(0)}
\mathbb{P}_{k'}\mathbb{P}_{\mu} \mathbb{P}_{k}) \; ,
\label{W11 1 b}
\end{eqnarray}
\label{W11 1}
\end{subequations}
where we have used Eq. (\ref{l(0)=1}) and we have defined
\begin{equation}
G_{kk'}(\epsilon_1) = \delta_{k, k'} + \lambda(\epsilon_1)(1-\delta_{k, k'}) .
\label{Gkk'}
\end{equation}

An important result is that Eq. (\ref{W11 1}) can be inverted to obtain $\rho_s^{(0)}$, giving \cite{johansen-mello:2008,amir-pier}
\begin{equation}
\langle k|  \rho_{s}^{(0)}  | k' \rangle
= \sum_{\mu}
\frac{W^{( \mathbb{P}_{\mu} \leftarrow \mathbb{P}_{k})}
_{1 1} (\epsilon_1)}
{G_{kk'}(\epsilon_1)} \cdot 
\frac{(\mu|k'\rangle}
{(\mu|k\rangle} \; ,
\label{rho elements from W(b,a) 1}
\end{equation}
It is clear that Eq. (\ref{rho elements from W(b,a) 1}) requires 
$(\mu|k\rangle \neq 0 \; \forall \mu, k$, i.e., that the two bases must be mutually non-orthogonal. 
Eq. (\ref{rho elements from W(b,a) 1}) is the main result of this chapter. 

As a consequence, we see that the set of complex quantities 
$W^{( \mathbb{P}_{\mu} \leftarrow \mathbb{P}_{k})}_{1 1} (\epsilon_1)$,
$\forall k,\mu$, 
contains all the information about the state of the system $\rho_s^{(0)}$.
If we could infer 
$W^{( \mathbb{P}_{\mu} \leftarrow \mathbb{P}_{k})}_{1 1} (\epsilon_1)$,
$\forall k,\mu$,
from measurement outcomes, we could reconstruct the full $\rho_s^{(0)}$.

Notice that both the real and imaginary parts of the {\em complex} quantities 
$W^{( \mathbb{P}_{\mu} \leftarrow \mathbb{P}_{k})}_{1 1} (\epsilon_1)$
are needed for tomography.
However, from the {\em detected} position-position correlations
$\langle \hat{Q}_1 \hat{Q}_2 \rangle
^{(\mathbb{P}_{\mu} \leftarrow \mathbb{P}_{k})}
/\epsilon_1 \epsilon_2$,
we directly extract only
$\Re W^{( \mathbb{P}_{\mu} \leftarrow \mathbb{P}_{k})}_{1 1} (\epsilon_1)$,
as we see from Eq. (\ref{<Q1Q2> P P b}).
In order to find 
$\Im W^{( \mathbb{P}_{\mu} \leftarrow \mathbb{P}_{k})}_{1 1} (\epsilon_1)$, 
we use the momentum-position correlation.
Our aim is to show that the measured quantities, the position-position and momentum-position correlation functions, are informationally complete: 
that is, one can reconstruct 
$W^{( \mathbb{P}_{\mu} \leftarrow \mathbb{P}_{k})}_{1 1}$ from these quantities.

We recall, from Eq. (\ref{<Q> for proj}), the analogous situation in the single-measurement case, where the measurable quantities
$\langle \hat{Q} \rangle_f^{(\hat{\mathbb{P}}_{a_{\nu}})}$
allow the reconstruction of the diagonal matrix elements of 
$ \rho_s^{(0)}$, i.e.,
$
(1/\epsilon)\langle \hat{Q} \rangle_f^{(\hat{\mathbb{P}}_{a_{\nu}})}
=
W_{\tau=1}^{(\hat{\mathbb{P}}_{a_{\nu}})}
=\langle a_{\nu} | \rho_s^{(0)} |a_{\nu} \rangle 
$.
The extension of this result to the two-probe case is 
Eq. (\ref{rho elements from W(b,a) 1}), in conjunction with Eq. (\ref{<Q1Q2> P P b})
(that relates 
$\Re W^{( \mathbb{P}_{\mu} \leftarrow \mathbb{P}_{k})}_{1 1} (\epsilon_1)$ 
to measurable quantities), and 
Eq. (\ref{y}) below (that relates 
$\Im W^{( \mathbb{P}_{\mu} \leftarrow \mathbb{P}_{k})}_{1 1} (\epsilon_1)$
to measurable quantities).

In the present case, Eq. (\ref{<P1Q2> 0}) for the probes momentum-position correlation function gives
\begin{equation}
\frac{1}{\epsilon_1 \epsilon_2}
\langle \hat{P}_1 \hat{Q}_2 \rangle
^{(
\mathbb{P}_{\mu} \leftarrow \mathbb{P}_{k}
)}
= 2 \sigma_{P_1}^2 \Im \widetilde{W}^{( \mathbb{P}_{\mu} \leftarrow \mathbb{P}_{k})}
_{1 1} (\epsilon_1),
\label{<P1Q2> P P}
\end{equation}
where 
\begin{equation}
\widetilde W^{( \mathbb{P}_{\mu} \leftarrow \mathbb{P}_{k})}_{1 1} (\epsilon_1)
= {\rm Tr} (\rho_{s}^{(0)}\mathbb{P}_{k}\mathbb{P}_{\mu} \mathbb{P}_{k})
+\tilde\lambda(\epsilon_1) \sum_{k'(\neq k)} {\rm Tr} (\rho_{s}^{(0)}
\mathbb{P}_{k'}\mathbb{P}_{\mu} \mathbb{P}_{k})\; .
\label{tildeW P P 0}
\end{equation}
Although the function $\widetilde{W}$ is in general not equal to the function $W$ (except when the probes are described by pure Gaussian states, as in Ref. \cite{johansen-mello:2008}), we now prove that it contains the necessary information to find the imaginary part of 
$W^{( \mathbb{P}_{\mu} \leftarrow \mathbb{P}_{k})}_{1 1}$, and therefore enables a complete state reconstruction. 

If we write 
\begin{subequations}
\begin{eqnarray}
W^{(\mathbb{P}_{\mu} \leftarrow \mathbb{P}_{k})}_{1 1} 
= x_{\mu k} + i y_{\mu k}, 
\label{x,y}  \\
\widetilde{W}^{(\mathbb{P}_{\mu} \leftarrow \mathbb{P}_{k})}_{1 1} 
= \tilde{x}_{\mu k} + i  \tilde{y}_{\mu k},
\label{xtilde,ytilde}
\end{eqnarray}
\label{x,y,xtilde,ytilde}
\end{subequations}
the correlation functions,
Eqs. (\ref{<Q1Q2> P P b}) and (\ref{<P1Q2> P P}), become
\begin{subequations}
\begin{eqnarray}
\frac
{\langle \hat{Q}_1 \hat{Q}_2 \rangle
^{(
\mathbb{P}_{\mu} \leftarrow \mathbb{P}_{k}
)}}
{\epsilon_1 \epsilon_2}
&=& x_{\mu k}, 
\label{<Q1Q2>=x}         \\
\frac
{\langle \hat{P}_1 \hat{Q}_2 \rangle
^{(
\mathbb{P}_{\mu} \leftarrow \mathbb{P}_{k}
)}}
{\epsilon_1 \epsilon_2}
&=& 2 \sigma_{P_1}^2 \tilde{y}_{\mu k}.
\label{<P1Q2>=ytilde}
\end{eqnarray}
\label{<Q1Q2>,<Q1P2>}
\end{subequations}

Our aim is to express $y_{\mu k}$ in terms of the measured quantities of
Eqs. (\ref{<Q1Q2>,<Q1P2>}).
We go back to the expressions (\ref{W11 1}) and (\ref{tildeW P P 0})
for 
$W^{( \mathbb{P}_{\mu} \leftarrow \mathbb{P}_{k})}_{1 1} (\epsilon_1)$ and
$\widetilde{W}^{( \mathbb{P}_{\mu} \leftarrow \mathbb{P}_{k})}_{1 1} (\epsilon_1)$.
The quantities $\lambda(\epsilon_1)$, $\tilde{\lambda}(\epsilon_1)$  are known if the state of the probe ${\pi}_1$ is known 
(see Eqs. (\ref{lambda,g,h}) and (\ref{tilde-bar-lambda})).
We write them as
\begin{subequations}
\begin{eqnarray}
\lambda(\epsilon_1) 
&=& \lambda_r(\epsilon_1) + i \lambda_i(\epsilon_1), 
\label{lambda r,i}        \\ 
\tilde{\lambda}(\epsilon_1) 
&=& \tilde{\lambda}_r(\epsilon_1) + i \tilde{\lambda}_i(\epsilon_1).
\label{lambda_tilde r,i}
\end{eqnarray}
\label{lambda r,i;lambda_tilde r,i}
\end{subequations}
On the other hand, the traces appearing in Eqs. (\ref{W11 1}) and 
(\ref{tildeW P P 0}) are unknown; we write them as
\begin{subequations}
\begin{eqnarray}
 {\rm Tr} (\rho_{s}^{(0)}\mathbb{P}_{k}\mathbb{P}_{\mu} \mathbb{P}_{k})
&=& 
|\langle k | \mu) |^2 
\langle k |\rho_s | k \rangle
= |\langle k | \mu)|^2
\sum_{\mu'} x_{\mu' k} 
\label{r0}
\\
\sum_{k'(\neq k)} {\rm Tr} (\rho_{s}^{(0)}
\mathbb{P}_{k'}\mathbb{P}_{\mu} \mathbb{P}_{k})
&=& r_{\mu k}+is_{\mu k} .
\label{r,s}
\end{eqnarray}
\label{r0,r,s}
\end{subequations}
Using Eq. (\ref{rho elements from W(b,a) 1}), we wrote Eq. (\ref{r0}) in terms of measured quantities only.

We introduce the definitions (\ref{x,y,xtilde,ytilde}), 
(\ref{lambda r,i;lambda_tilde r,i}), (\ref{r0,r,s}) in Eqs.~(\ref{W11 1}) and (\ref{tildeW P P 0}), which then give
\begin{subequations}
\begin{eqnarray}
x_{\mu k}
&=& |\langle k | \mu)|^2
\sum_{\mu'} x_{\mu' k} + \lambda_r r_{\mu k} -\lambda_i  s_{\mu k} ,
\\
y_{\mu k}
&=& \lambda_i 
r_{\mu k}  
+ \lambda_r
s_{\mu k}, 
   \\
\tilde{y}_{\mu k}
&=& \tilde{\lambda}_i 
r_{\mu k}  
+  \tilde{\lambda}_r
s_{\mu k}.
\end{eqnarray}
\end{subequations}
For every pair of indices $\mu, k$ we now have a system of three linear equations in the three unknowns $r_{\mu k}$, $s_{\mu k}$ and $y_{\mu k}$, which can thus be expressed in terms of the measured quantities 
$\tilde{y}_{\mu k}$ and the $x_{m k}$
of Eqs. (\ref{<Q1Q2>,<Q1P2>}).
The result for $y_{\mu k}$ is
\begin{equation}
y_{\mu k}
= \frac{\Im \{\lambda(\epsilon_1)\tilde\lambda^\ast(\epsilon_1)\}}
{\Re \{\lambda(\epsilon_1)\tilde\lambda^\ast(\epsilon_1)\}}
\;\Big(x_{\mu k}-|\langle k|\mu\rangle|^2 
\sum_{\mu'} x_{\mu' k}\Big)
+\frac{|\lambda(\epsilon_1)|^2}{\Re \{\lambda(\epsilon_1)\tilde\lambda^\ast(\epsilon_1)\}}\;\tilde{y}_{\mu k} .
\label{y}
\end{equation}

We have thus achieved our goal of expressing 
$W^{( \mathbb{P}_{\mu} \leftarrow \mathbb{P}_{k})}_{1 1} (\epsilon_1)$, 
and hence  $\rho_s^{(0)}$ of Eq.~(\ref{rho elements from W(b,a) 1}),
in terms of the measured correlations of Eqs.~(\ref{<Q1Q2>,<Q1P2>}).
This completes our procedure.

It is interesting to examine the strong- and weak-coupling limits of the above procedure.
In the strong-coupling limit, $\epsilon_1 \to \infty$, from Eq. (\ref{Gkk'}) we see that
$G_{k'k}(\epsilon_1) \to \delta_{k' k}$,
and Eq. (\ref{rho elements from W(b,a) 1}) gives
\begin{equation}
\langle k|  \rho_{s}^{(0)}   | k \rangle
\to
\sum_{\mu} \;
{\rm Tr}\left(\rho_s^{(0)}
\mathbb{P}_{k}
\mathbb{P}_{\mu}
\mathbb{P}_{k}
\right)
= \sum_{\mu}\;  
{\cal W}_{\mu k}^{(\mathbb{P}_{\mu}\leftarrow \mathbb{P}_{k})}
= {\rm Tr}_s (\rho_s^{(0)} \mathbb{P}_{k} ) \; ,
\label{rho retrieval strong coupling}
\end{equation}
in terms of Wigner's joint probability defined in Eq. (\ref{W e to infty}).
Notice that in this limit only the diagonal elements $\rho_s^{(0)}$  can be retrieved.
This is the limit in which Wigner's formula (\ref{W e to infty}) arises.

In the weak-coupling limit,
$\epsilon_1 \to 0$, we have $G_{k'k}(\epsilon_1) \to 1$, and 
Eq. (\ref{rho elements from W(b,a) 1}) gives
\begin{equation}
\langle k|  \rho_{s}^{(0)}  | k' \rangle
\to
\sum_{\mu} 
{\rm Tr}_s (\rho _s^{(0)} \mathbb{P}_{\mu}\mathbb{P}_{k})
  \frac{(\mu|k'\rangle}
  {(\mu|k\rangle} 
=\sum_{\mu} 
{\cal K}_{\mu k}^{(\mathbb{P}_{\mu}\leftarrow \mathbb{P}_{k})}
  \frac{(\mu|k'\rangle}
  {(\mu|k\rangle},
  \label{rho retrieval weak coupling}
\end{equation}
in terms of Kirkwood's joint quasi-probability defined in Eq. (\ref{W e to 0}).
The result (\ref{rho retrieval weak coupling}) was first obtained in 
Ref. \cite{Johansen-Quantheosuccproj:07}.

At first glance it seems that, in a $N$-dimensional Hilbert space, the present scheme for state reconstruction requires the measurement of the $2N^2$ different correlations
$\langle \hat{Q}_1 \hat{Q}_2 \rangle^{(\mathbb{P}_{\mu}\leftarrow \mathbb{P}_{k})}$
and 
$\langle \hat{P}_1 \hat{Q}_2 \rangle^{(\mathbb{P}_{\mu}\leftarrow \mathbb{P}_{k})}$.
However, Hermiticity and the unit value of the trace of the density matrix $\rho_s^{(0)}$ impose $N^2+1$ restrictions among its matrix elements, so that $\rho_s^{(0)}$
can be expressed in terms of $N^2-1$ independent parameters. 
These restrictions eventually imply that only $N^2-1$ of these correlations are actually independent and thus 
the measurement of only $N^2-1$ correlations is required.

Let us take as an example $N=2$. Labelling the states as $k=0,1$ and $\mu=+,-$,
it is enough to measure the correlations
\begin{equation}
\langle \hat{Q}_1 \hat{Q}_2 \rangle
^{(
\mathbb{P}_{+} \leftarrow \mathbb{P}_{0}
)},
\hspace{2mm}
\langle \hat{Q}_1 \hat{Q}_2 \rangle
^{(
\mathbb{P}_{-} \leftarrow \mathbb{P}_{0} 
)}
,
\hspace{2mm} {\rm and} \hspace{2mm}
\langle \hat{P}_1 \hat{Q}_2 \rangle
^{(
\mathbb{P}_{-} \leftarrow \mathbb{P}_{0}
)} \; .
\label{correlations needed for N=2}
\end{equation}
In terms of these correlations and using the notation of 
Eqs. (\ref{<Q1Q2>,<Q1P2>}), the density matrix elements of our system prior to the mesasurement are given by
\begin{subequations}
\begin{eqnarray}
\rho_{00}^{(0)} &=& x_{+,0} +  x_{-,0}   \\
\rho_{11}^{(0)} &=& 1 - x_{+,0} -  x_{-,0} \\
\rho_{01}^{(0)} &=& \frac{x_{+,0} -  x_{-,0} - 2i y_{-,0}}{g(\epsilon_1)} \\
\rho_{10}^{(0)} &=& \frac{x_{+,0} -  x_{-,0} + 2i y_{-,0}}{g(\epsilon_1)} \; .
\end{eqnarray}
\label{rho in terms of the x and y correlations N=2 Gaussian}
\end{subequations}
We have used Gaussian states for the probes, for which 
$y_{\mu k}=\tilde{y}_{\mu k}$ (see Eqs. (\ref{g,gtilde,gaussian case})).

The relations appearing in Eqs. (\ref{W11 1}) between 
$W^{( \mathbb{P}_{\mu} \leftarrow \mathbb{P}_{k})}_{11} (\epsilon_1)$
and $\rho_s^{(0)}$
shed light on the the strong- and weak-couping limits of the retrieval scheme described above.
We consider again the case $N=2$, for Gaussian states for the probes, and for the case in which the states $|k\rangle$, $k=0,1$, are eigenstates of the Pauli matrix $\sigma_k$ and the states $|\mu )$, $\mu=+,-$, are eigenstates of $\sigma_x$.

In the strong-coupling limit $\epsilon_1 \to \infty$,
\begin{equation}
W^{( \mathbb{P}_{\mu} \leftarrow \mathbb{P}_{k})}
_{11} (\epsilon_1)
\to
{\rm Tr}\left(\rho_s^{(0)}
\mathbb{P}_{k}
\mathbb{P}_{\mu}
\mathbb{P}_{k}
\right)
=|\langle k|\mu)|^2 \rho_{kk}^{(0)},
\end{equation}
which, for $N=2$ give
\begin{subequations}
\begin{eqnarray}
W^{(+\leftarrow 0)}_{11} (\epsilon_1) \to \frac12 \rho^{(0)}_{00} , \;\;\;
W^{(-\leftarrow 0)}_{11} (\epsilon_1) \to \frac12 \rho^{(0)}_{00} \\
W^{(+\leftarrow 1)}_{11} (\epsilon_1) \to \frac12 \rho^{(0)}_{11} , \;\;\;
W^{(-\leftarrow 1)}_{11} (\epsilon_1) \to \frac12 \rho^{(0)}_{11} \; .
\end{eqnarray}
\end{subequations}
Thus,
$W^{( \mathbb{P}_{\mu} \leftarrow \mathbb{P}_{k})}_{11} (\epsilon_1)$
only contains information on the diagonal elements of $\rho_s^{(0)}$, so that only $\rho^{(0)}_{00}$ and $\rho^{(0)}_{11}$ can be retrieved.

In the weak-coupling limit $\epsilon_1 \to 0$,
\begin{equation}
W^{( \mathbb{P}_{\mu} \leftarrow \mathbb{P}_{k})}
_{11} (\epsilon_1)
\to
{\rm Tr}\left( \rho_s^{(0)}\mathbb{P}_{\mu} \mathbb{P}_{k} \right)
={\cal K}_{\mu k}
=\langle k|\rho_s^{(0)}|\mu)(\mu |k\rangle \;,  
\end{equation}
and for $N=2$
\begin{subequations}
\begin{eqnarray}
W^{(+\leftarrow 0)}_{11} (\epsilon_1) 
&\to& 
\frac{\rho^{(0)}_{00} +\rho^{(0)}_{01} }{2} , \;\;\;\;
W^{(-\leftarrow 0)}_{11} (\epsilon_1) \to  
\frac{\rho^{(0)}_{00} -\rho^{(0)}_{01} }{2} \\
W^{(+\leftarrow 1)}_{11} (\epsilon_1) 
&\to&   
\frac{\rho^{(0)}_{10} +\rho^{(0)}_{11} }{2} ,  \;\;\;\;
W^{(-\leftarrow 1)}_{11} (\epsilon_1) \to 
\frac{\rho^{(0)}_{10} -\rho^{(0)}_{11}}{2} \; . 
\end{eqnarray}
\end{subequations}
We thus see that the four $\rho_s^{(0)}$ matrix elements can be retrieved.

In conclusion, to reconstruct a QM state using the successive-measurement scheme, it is better to perform  measurements with weak coupling $\epsilon_1$, rather than with strong coupling.

For the case of an arbitrary $\epsilon_1$ we have the relations
(\ref{rho in terms of the x and y correlations N=2 Gaussian})
giving the $\rho_s^{(0)}$ matrix elements in terms of the position-position and momentum-position correlations.
Recall that 
$
g(\epsilon_1)=\exp{\left(-\frac{\epsilon_1^2}{8 \sigma_{Q_1}^2}\right)}
$.
Even if $g\neq 0$, but $g \ll 1$, a small experimental uncertainty in extracting
$\langle Q_1 Q_2\rangle^{(\mu \leftarrow k)}$ and 
$\langle P_1 Q_2\rangle^{(\mu \leftarrow k)}$, which give
$W^{(\mu \leftarrow k)}_{1 1} (\epsilon_1)
= x_{\mu, k}+ iy_{\mu, k}$, 
is divided by a small number $g \ll 1$
when $k \neq k'$, and this makes the error in extracting $\rho^{(0)}_{kk'}$
large.
Again, we see that, in general, it is advantageous to use a weak coupling rather than a strong coupling.

\section{A quasi-distribution and a generalized transform of observables}
\label{generalized transform}

Conceptually, one attractive feature of the tomographic approach we have described
is that the quantities 
$W^{( \mathbb{P}_{\mu} \leftarrow \mathbb{P}_{k})}_{1 1} (\epsilon_1)$
that enter the resonstruction formula, Eq. (\ref{rho elements from W(b,a) 1}),
can be interpreted as a {\em quasi-probability}, as we now explain.

Let $\hat{O}$ be an observable associated with an $N$-dimensional quantum system.
Using Eq. (\ref{rho elements from W(b,a) 1}), we can express its expectation value as
\begin{subequations}
\begin{eqnarray}
{\rm Tr}_s(\hat{\rho}_s^{(0)}\hat{O})
= \sum_{k k'} \langle k| \rho_s^{(0)} |k'\rangle \langle k'| \hat{O} |k\rangle
= \sum_{k \mu}
W^{( \mathbb{P}_{\mu} \leftarrow \mathbb{P}_{k})}_{1 1} (\epsilon_1) \;
O(\mu, k; \epsilon_1) \; ,
\label{<O> (W11,O(mu,k)) a}
\end{eqnarray}
where we have defined the ``transform" of the operator $\hat{O}$ as 
\begin{eqnarray}
O(\mu, k; \epsilon_1)
=\sum_{k'}
\frac{(\mu|k'\rangle}
  {(\mu|k\rangle}
\frac{\langle k' |  \hat{O} | k \rangle}
{G_{k' k}(\epsilon_1)} \; .
\label{O(mu,k)}
\end{eqnarray}
\label{<O> (W11,O(mu,k))}
\end{subequations}
Eqs. (\ref{<O> (W11,O(mu,k))}) have a structure similar to that of a number of transforms found in the literature, that express the quantum mechanical expectation value of an observable in terms of its transform and a quasi-probability distribution.

For example, the Wigner transform of an observable and the Wigner function of a state are defined in the phase space $(q,p)$ of the system, $q$ and $p$ labelling 
the states of the coordinate and momentum bases, respectively. 

In the present case, the transform (\ref{O(mu,k)}) of the observable is defined for the pair of variables $(\mu, k)$, $\mu$ and $k$ labelling the states of 
each of the two bases.
As Eq. (\ref{<O> (W11,O(mu,k)) a}) shows, the quantity 
$W^{( \mathbb{P}_{\mu} \leftarrow \mathbb{P}_{k})}_{1 1} (\epsilon_1)$ 
plays the role of a {\em quasi-probability} for the system state
$\hat{\rho}_s^{(0)}$, and is also defined for the pair of variables 
$(\mu, k)$.
It can be thought of as the joint quasi-probability (in general complex \cite{feynmann87}) of two non-degenerate observables, the two bases being their respective eigenbases. 
Since any pair of mutually orthogonal bases can be used, we have a whole family of transforms that can be employed to retrieve the state.

In the literature it has been discussed how Wigner's function can be considered as a representation of a quantum state 
(see, e.g., Ref. \cite{schleich-2001}, Chs. 3 and 4), in the sense that i) it allows retrieving the density operator, and ii) any quantum-mechanical expectation value can be evaluated from it.
Similarly, and for the same reasons, in the present context the quasi-probability 
$W^{( \mathbb{P}_{\mu} \leftarrow \mathbb{P}_{k})}_{1 1} (\epsilon_1)$
can also be considered as a representation of a quantum state.

\section{Conclusions}
\label{concl}

In this series of lectures we described how to obtain information on a quantum-mechanical system, by coupling it to an auxiliary degeree of freedon, or probe, which is then detected by some measuring instrument. 
The model used in the analysis is the one introduced by von Neumann, in which the interaction of the system proper with the probe is described in a dynamical way.

For single measurements we used the standard von Neumann model, which employs one probe coupled to the system with an arbitrary coupling strength; 
for successive measurements, we generalized von Neumann's model employing two probes.

In the case of single measurements, we investigated the average, the variance and the full distribution of the probe position after the interaction with the system, and their relation with the properties of the latter.
An interesting outcome of the analysis is the reduced density operator of the system which, in the limit of strong coupling between the system and the probe, was shown to reduce to the von-Neumann-L\"uders rule, a result which is frequently obtained in a ``non-dynamical" way, as a result of non-selective projective measurements.

In the case of successive measurements, we studied how to obtain information on the system by detecting the position-position and momentum-position correlations of the two probes.
We saw that the so-called ``Wigner's formula", as well as 
``Kirkwood's quasi-probability distribution", emerge in the strong- and in the weak-coupling limits, respectively, of the above formalism.
We investigated the successive measurement of the same observable and showed how, in the strong-coupling limit, the result behaves in the expected manner.
The relation of the weak-value theory to successive measurements was briefly mentioned.

Furthermore, we described a quantum state tomography scheme which is applicable to a system described in a Hilbert space of arbitrary finite dimensionality, which is constructed from sequences of two measurements.
The scheme consists of measuring the various pairs of projectors onto two bases --which have no mutually orthogonal vectors.

Finally, we found a generalized transform of the state and the observables 
based on the notion of successive measurements.
The result has a structure similar to that of a number of transforms found in the literature, like the Wigner function, that express the quantum-mechanical expectation value of an observable in terms of its transform and a quasi-probability distribution.

In a recent investigation, the question was posed whether it is possible to find appropriate measurements of the system position and momentum that would allow the reconstruction of the Wigner function of the system state.
It was found that the types of measurements needed are successive measurements of projectors associated with position and momentum, of the type envisaged by von Neumann's model which was discussed here.
A preliminary account of that investigation can be found in 
Ref. \cite{mello-revzen_2013}.





\begin{theacknowledgments}

The author acknowledges financial support from Conacyt, Mexico (under Contract No. 79501) and from the Sistema Nacional de Investigadores, Mexico.

\end{theacknowledgments}



\bibliographystyle{aipproc}   

\bibliography{mello}

\IfFileExists{\jobname.bbl}{}
 {\typeout{}
  \typeout{******************************************}
  \typeout{** Please run "bibtex \jobname" to optain}
  \typeout{** the bibliography and then re-run LaTeX}
  \typeout{** twice to fix the references!}
  \typeout{******************************************}
  \typeout{}
 }

\end{document}